\documentclass[12pt,pre,aps,showpacs,preprint,nofootinbib]{revtex4}

\usepackage{psfig}
\usepackage{amsfonts}
\begin{document}

\title{Random Sequential Addition of Hard Spheres in High Euclidean Dimensions}

\author{S. Torquato}

\email{torquato@electron.princeton.edu}

\affiliation{\emph{Department of Chemistry}, \emph{Princeton University}, Princeton
NJ 08544}

\affiliation{\emph{Program in Applied and Computational Mathematics}, \emph{Princeton
University}, Princeton NJ 08544}

\affiliation{\emph{PRISM, Princeton University}, Princeton NJ 08544}

\affiliation{\emph{Princeton Center for Theoretical Physics, Princeton University}, Princeton
NJ 08544}

\author{O. U. Uche}

\affiliation{\emph{Department of Chemical Engineering}, \emph{Princeton University}, Princeton
NJ 08544}

\author{F. H. Stillinger}

\affiliation{\emph{Department of Chemistry}, \emph{Princeton University}, Princeton
NJ 08544}

\begin{abstract}
Sphere packings in high dimensions have been the subject of recent theoretical
interest. Employing numerical and theoretical
methods, we investigate the structural characteristics  of random sequential
addition (RSA)  of congruent spheres in $d$-dimensional Euclidean space 
$\mathbb{R}^d$ in the infinite-time or saturation limit for the first six
space dimensions ($1 \le d \le 6$). Specifically, we determine the saturation 
density, pair correlation function, cumulative coordination number and the structure factor in each
of these dimensions. We find that for $2 \le d  \le 6$, the saturation density $\phi_s$ scales 
with dimension as $\phi_s= c_1/2^d+c_2 d/2^d$, where $c_1=0.202048$ and $c_2=0.973872$.
We also show analytically that the same density scaling persists in the high-dimensional limit,
albeit with different coefficients. A byproduct of this high-dimensional analysis is a 
relatively sharp lower bound on the saturation density for any $d$ given by  
$\phi_s \ge (d+2)(1-S_0)/2^{d+1}$, where $S_0\in [0,1]$ is the structure factor at $k=0$ 
(i.e., infinite-wavelength number variance) in the high-dimensional
limit. We prove rigorously that a Pal{\`a}sti-like conjecture
(the saturation density in $\mathbb{R}^d$ is equal to that of the one-dimensional problem 
raised to the $d$th power) cannot be true for RSA hyperspheres. We demonstrate that the structure
factor $S(k)$  must be analytic at $k=0$ and that RSA packings for $1 \le d \le 6$
are nearly ``hyperuniform." Consistent with the recent ``decorrelation principle," we find
that pair correlations markedly diminish as the space dimension increases up to six.
We also obtain kissing (contact)
number statistics for saturated RSA configurations on the surface
of a $d$-dimensional sphere for dimensions $2 \le d \le 5$
and compare to the maximal kissing numbers in these dimensions.
We determine the structure factor exactly for the related ``ghost"
RSA packing in $\mathbb{R}^d$ and show that its distance from ``hyperuniformity"
increases as the space dimension increases, approaching a constant
asymptotic value of $1/2$.
Our work has implications for the possible
existence of disordered classical ground states for some continuous potentials
in sufficiently high dimensions.

\end{abstract}

\maketitle

\section{Introduction}

We call a collection of congruent spheres in $d$-dimensional Euclidean space $\mathbb{R}^d$ 
a hard-sphere packing if no two spheres overlap.
The study of the structure and macroscopic properties of hard spheres in physical
dimensions ($d=2$ or $3$) has a rich history, dating back to at least the work of Boltzmann \cite{Bo64}.
Hard-sphere packings have been used to model a variety of systems, including liquids \cite{Ha86},
amorphous and granular media \cite{To02a}, and crystals \cite{Chaik95}. 
There has been resurgent interest in hard-sphere packings in dimensions
greater than three in both the physical and mathematical sciences. For example, it is known 
that the optimal way of sending digital signals over noisy channels corresponds 
to the densest sphere packing in a high dimensional space \cite{Co93}.
These ``error-correcting" codes underlie a variety of systems in digital
communications and storage, including compact disks, cell phones and the Internet.
Physicists have studied hard-sphere packings in high dimensions to gain insight
into ground and glassy states of matter as well as 
phase behavior in lower dimensions \cite{Fr99,Pa00,To02c,Pa06}. 

The determination of the densest packings
in arbitrary dimension is a problem of long-standing interest in discrete geometry \cite{Co93}.
The {\it packing density} or simply density $\phi$ of a sphere packing is the fraction of
space $\mathbb{R}^d$ covered by the spheres, i.e.,
\begin{equation}
\phi=\rho v_1(R),
\label{phi}
\end{equation}
where $\rho$ is the number density,
\begin{equation}
v_1(R) = \frac{\pi^{d/2}}{\Gamma(1+d/2)} R^d
\label{v(R)}
\end{equation}
is the volume of a $d$-dimensional sphere of radius $R$, and
$\Gamma(x)$ is the gamma function. We call 
\begin{equation}
\phi_{\mbox{\scriptsize max}}= \sup_{P\subset \mathbb{R}^d} \phi(P)
\end{equation}
the {\it maximal density}, where the supremum is taken over
all packings in $\mathbb{R}^d$.
The sphere packing problem seeks to answer the following  
question: Among all packings of congruent spheres,
what is the maximal packing density $\phi_{\mbox{\scriptsize max}}$, i.e., largest
fraction of $\mathbb{R}^d$ covered by the spheres,
and what are the corresponding arrangements of the spheres \cite{Ro64,Co93}?
For $d=1,2$ and $3$, the optimal solutions are known. For $3< d <10$,
the densest known packings of congruent spheres are Bravais lattice
packings \cite{Co93}, but in sufficiently large dimensions
the optimal packings are likely to be non-Bravais lattice packings.
Upper and lower bounds on the maximal density
$\phi_{\mbox{\scriptsize max}}$
exist in all dimensions \cite{Co93}. For example, Minkowski \cite{Mi05} proved that the maximal density 
$\phi^L_{\mbox{\scriptsize max}}$ among all Bravais lattice packings 
for $d \ge 2$ satisfies the lower bound
\begin{equation}
\phi^L_{\mbox{\scriptsize max}} \ge \frac{\zeta(d)}{2^{d-1}},
\label{mink}
\end{equation}
where $\zeta(d)=\sum_{k=1}^\infty k^{-d}$ is the Riemann zeta function.
It is seen that for large values of $d$,
the asymptotic behavior of the {\it nonconstructive} Minkowski lower bound is controlled by $2^{-d}$.
Note that the density of a {\it saturated} packing of congruent spheres
in $\mathbb{R}^d$ for all $d$ satisfies 
\begin{equation}
\phi \ge \frac{1}{2^d},
\label{sat}
\end{equation}
which has the same dominant exponential term as (\ref{mink}).
A saturated packing of congruent spheres
of unit diameter  and density $\phi$ in $\Re^d$ has the property that each point in space lies
within a unit distance from the center of some sphere.  
In the large-dimensional limit, Kabatiansky and Levenshtein   \cite{Ka78} showed that the maximal density
is bounded from above according to the asymptotic upper bound
\begin{equation}
\phi_{\mbox{\scriptsize max}} \le \frac{1}{2^{0.5990\,d}}.
\label{kab}
\end{equation}

The present paper is motivated by some recent work on disordered sphere-packings
in high dimensions \cite{To06a,To06b}.
In Ref. \cite{To06a}, we introduced a generalization of the well-known random sequential addition (RSA) process
for hard spheres in $d$-dimensional Euclidean space $\mathbb{R}^d$. 
This model can be viewed as a special ``thinning" of a Poisson point process such that the subset of points
at the end of the thinning process corresponds to a sphere packing. One obvious rule is to retain a test
sphere at time $t$ only if it does not overlap a sphere that was successfully
added to the packing at an earlier time. This criterion defines the standard RSA process in $\mathbb{R}^d$
\cite{To02a,Ta00}, which generates a homogeneous and isotropic
sphere packing in $\mathbb{R}^d$ with a time-dependent density
$\phi(t)$. In the limit $t \rightarrow \infty$, the RSA process corresponds to a saturated
packing with a maximal or {\it saturation} density $\phi(\infty) \equiv \lim_{t\rightarrow
\infty} \phi(t)$ \cite{footnote1}. In one dimension, the RSA process is commonly known as the ``car parking problem", 
which Re{\' n}yi showed has a saturation density $\phi(\infty)= 0.747598\ldots$ \cite{Re63}.
For $2 \le d < \infty$, an exact determination of $\phi(\infty)$ is not
possible, but estimates for it have been obtained via computer experiments
in two dimensions (circular disks) \cite{Fe80,Hi86} and three dimensions (spheres) \cite{Co88,Ta91}.
However, estimates of the saturation density $\phi(\infty)$ in 
higher dimensions have heretofore not been obtained.

Another thinning criterion retains a test sphere centered at position $\bf r$ at time $t$ 
if no other test sphere is within a unit radial distance from $\bf r$ for
the time interval $\kappa t$ prior to $t$, where $\kappa$ is a positive constant
in the interval $[0,1]$. This packing is a subset of the RSA packing, which
we call the generalized RSA process. Note
that when $\kappa=0$, the standard RSA process is recovered, and when $\kappa=1$,
we obtain the ``ghost" RSA process \cite{To06a}, which is amenable
to exact analysis. In particular, we showed that the $n$-particle correlation function  
$g_n({\bf r}_1,{\bf r}_2,\dots,{\bf r}_n)$  for the ghost RSA  packing can be obtained analytically for any $n$,
all allowable densities and in any dimension. This represents the first exactly
solvable disordered sphere-packing model in arbitrary dimension. 
For statistically homogeneous  packings in $\mathbb{R}^d$, these correlation functions
are defined so that $\rho^n g_n({\bf r}_1,{\bf r}_2,\dots,{\bf r}_n)$  is proportional to
the probability density for simultaneously finding $n$ particles at
locations ${\bf r}_1,{\bf r}_2,\dots,{\bf r}_n$ within the system,
where $\rho$ is the number density. Thus, in a packing without long-range order, each $g_n$ approaches
unity when all particle positions become widely separated within $\mathbb{R}^d$,
indicating no spatial correlations.
The fact that the maximal density $\phi(\infty)=1/2^d$ of the ghost RSA  packing implies
that there may be disordered sphere packings in sufficiently high $d$ whose density exceeds
Minkowski's lower bound  (\ref{mink}).

Indeed, in Ref. \cite{To06b}, a conjectural lower bound on the density of 
disordered sphere packings \cite{footnote2} was employed to provide the putative
exponential improvement on Minkowski's 100-year-old bound. The asymptotic behavior 
of the conjectural lower bound is controlled by $2^{-(0.77865\ldots) d}$.
These results suggest that the densest packings in sufficiently
high dimensions may be disordered rather than periodic, implying
the existence of disordered classical ground states for some continuous potentials.
In addition, a {\it decorrelation} principle for disordered packings was identified \cite{To06b},
which states that {\it unconstrained} correlations in disordered sphere packings 
vanish asymptotically in high dimensions
and that the $g_n$ for any $n \ge 3$ can be inferred entirely from a knowledge
of the number density $\rho$ and the pair correlation function $g_2({\bf r})$. At first glace, one might
be tempted to conclude that the decorrelation principle
is an expected  ``mean-field" behavior, which is not the case.  For example, it
is well known that in some spin systems correlations vanish
in the limit $d \rightarrow \infty$ and the system approaches the mean-field
behavior. While this notion is meaningful for spin
systems with attractive interactions, it is not for hard-core systems. The
latter is characterized by a total potential energy that is either zero or infinite, 
and thus cannot be characterized by a mean field. Furthermore, mean-field theories are limited to 
equilibrium considerations, and thus do not distinguish 
between ``constrained" and ``unconstrained" correlations that arise in non-equilibrium packings of which there
an infinite number of distinct ensembles. The decorrelation
principle is a statement about any disordered packing, equilibrium or not. 
For example, contact delta functions (constrained correlations) are
an important attribute of non-equilibrium jammed disordered packings and have no analog in equilibrium lattice
models of any dimension. Finally, the decorrelation principle 
arises from the fact that the Kabatiansky-Levenshtein asymptotic upper bound on the maximal packing 
density (\ref{kab}) implies that $\phi$ must go to zero at least as fast as $2^{-0.5990d}$ for large $d$
and therefore, unconstrained spatial correlations between spheres are expected
to vanish, i.e., statistical independence is established \cite{To06b}.
There is no counterpart of the Kabatiansky-Levenshtein bound in mean-field theories.

Motivated by these recent results, we study the structural properties of 
standard RSA packings ($\kappa=0$ for the generalized RSA process) in
the infinite-time or saturation limit, generated via computer simulations, for the first six
Euclidean space dimensions ($1\le d \le 6$). The algorithm is checked by reproducing
some known results for $d=1,2$ and $3$ \cite{Re63,Fe80,Hi86,Co88,Ta91}. Although we know that the saturation
density $\phi(\infty)$ is bounded from below by $2^{-d}$ \cite{To06a,To06b},
the manner in which  $\phi(\infty)$ scales with density is not known for
$d>3$. One objective of this paper is to answer this question. Another
aim is determine the corresponding pair correlation functions
and structure factors in order to ascertain whether decorrelations can be observed  
as the space dimension increases up to six. A byproduct of our high-dimensional analysis is a
relatively sharp lower bound on the saturation density of RSA packings for any $d$.
Although a Pal{\`a}sti-like conjecture
 (the saturation density in $\mathbb{R}^d$ is equal to that of the one-dimensional problem raised to the $d$th power)
is exact for ghost RSA packings, we rigorously prove that this conjecture
cannot be true for standard RSA packings.

In Appendix A, we obtain kissing (contact)
number statistics for saturated RSA configurations on the surface
of a $d$-dimensional sphere for dimensions $2 \le d \le 5$
and compare to the maximal kissing numbers in these dimensions.
In Appendix B, we determine the structure factor exactly for ghost
RSA packings and show that its distance from ``hyperuniformity" \cite{To03}
increases as the space dimension increases, approaching a constant
asymptotic value.

\section{Some Known Asymptotic Results for RSA Packings}

Here we collect some known asymptotic results for the standard
RSA process for $d$-dimensional hard spheres. Henceforth,
we call $\phi_s \equiv \phi(\infty)$ the saturation (infinite-time)
limit of the density.

In his numerical study of RSA hard disks,  Feder \cite{Fe80}
postulated that the asymptotic coverage in the long-time
limit for $d$-dimensional hard spheres follows the algebraic behavior
\begin{equation}
\phi_s-\phi(\tau) \sim \tau^{-1/d},
\label{Feder}
\end{equation}
where $\tau$ represents a dimensionless time.
Theoretical arguments supporting Feder's law (\ref{Feder})
have been put forth by Pomeau \cite{Po80} and  Swendsen \cite{Sw81}.
Not surprisingly, the saturation limit is approached
more slowly as the space dimension increases.

 Moreover, similar arguments lead to the conclusion
that the pair correlation function $g_{2}(r)$
at the saturation limit possesses a
 logarithmic singularity as the dimensionless radial
 distance $r$ for spheres of diameter $D$ approaches
 the contact value, independent of
dimension \cite{Po80,Sw81}, i.e.,
\begin{equation}
g_{2}(r) \sim \ln (r-D), \quad r\rightarrow D \; \mbox{and} \; \phi=\phi_s.
\label{rsa-contact}
\end{equation}
Boyer {\it et al.} \cite{Bo94} also showed that the pair correlation function
for $d=1$ has {\it super-exponential} decay. Specifically, they found
that at any finite time $\tau$ or density $\phi$,
\begin{equation}
g_2(r) \sim \frac{1}{\Gamma(r)}\left(\frac{2}{\ln (r-D)}\right)^{r-D},
\quad r\rightarrow \infty \; \mbox{and} \; 0<\phi \le \phi_s.
\end{equation}
Thus, $g_2$ is a short-ranged function at any density. This super-exponential
decay of the pair correlation function persists in higher dimensions as well.
As we will discuss in Section IV, this rapid decay of $g_2(r)$ has implications
for the analytic properties of the structure factor $S(k)$.

\section{Numerical Procedures}

In what follows, we describe
an efficient procedure to generate RSA packings in the saturation
limit as well as the methods used to compute structural
information, such as the density, pair correlation function, structure factor
and cumulative coordination number.

\subsection{Generation of RSA Packings in $\mathbb{R}^d$ in the Saturation Limit}

\label{sec:SL-RSA}

\noindent
We present a computationally fast method to generate RSA configurations of hard spheres
in the saturation limit in the thermodynamic limit. Periodic boundary conditions 
are applied to a hypercubic fundamental
cell of side length $L$ and volume $L^d$.
Spheres of diameter $D$ are placed randomly and sequentially inside the
fundamental cell, which is periodically replicated to fill all of 
$d$-dimensional Euclidean space $\mathbb{R}^d$, until the saturation limit
is achieved. 

In order to speed up the computation, we attempt to
add a particle only in the available space rather than wasting
computational time in attempting to add particles anywhere
in the fundamental cell \cite{Bro91,footnote3}. This requires keeping track
of the time-dependent available space, which is the 
space exterior to the union of the {\it exclusion} spheres
of radius $D$ centered at each successfully added sphere at any particular time.
This is done by tessellating the 
hypercubic fundamental cell into smaller, disjoint hypercubic ``voxels," which have
side length between $0.025 D$ and $0.1D$, 
depending upon the dimension. 

At the start of the simulation, all voxels are declared accessible to particle placement.
There are two stages involved to determine the time-dependent
available space. A coarse estimation of the available space
is used in the first stage, which is refined in the second
stage. In the first stage, a particle is successfully added to the simulation
box, provided that it does not overlap any existing particle.
To avoid checking for nonoverlaps with every successfully
added particle to the simulation box, we employ a ``neighbor list" ~\cite{To02a},
which amounts to checking within a local neighborhood of the
attempted particle placement. For each successfully added particle,
all voxels located within the largest
possible inscribed hypercube centered at the exclusion sphere of radius  $D$ that
are fully occupied by the particle are
declared to be part of the unavailable space. The voxels outside
this hypercube but within the exclusion sphere
may be partially filled. In the initial stages, such partially filled voxels are declared
to be part of the available space. We call these accessible voxels. If there have been at least one million unsuccessful placement
attempts since the last accepted particle placement, we move to the second 
stage to refine our determination of the available space.
In particular, we determine whether
each remaining accessible voxel from the first stage can accommodate a particle center
by a random search of each accessible voxel. After about 1000 random placement attempts, a particle
is either added to a particular voxel or this voxel is declared to be part of the unavailable space.
This search is carried out for all other accessible voxels.
The simulation terminates when all voxels are unavailable
for particle addition in the second stage.

This two-stage procedure enables us to generate RSA packings that
are saturated or nearly saturated. Generating truly saturated
RSA packings becomes increasingly difficult as the space dimension
increases, as the asymptotic relation (\ref{Feder}) indicates.

\subsection{Calculation of the Saturation Density}

At any instant of time $\tau$, the number $N(\tau)$ 
of added particles for a particular configuration
is known and the density  $\phi(\tau)$
is computed from relation (\ref{phi}) with $\rho=N(\tau)/L^d$.
We call $\phi_{stop} \equiv \phi(\tau_{max})$ the ``stopping" density, i.e., 
the density at the time $\tau_{max}$ when the simulation is terminated.
The system size $L/D$ is sufficiently large so as produce a histogram for 
$\phi_{max}$ or $\phi_s$ that is Gaussian distributed. Although we do not present
the full distribution of densities here, we do report the associated standard errors.
In order to estimate the true saturation density $\phi_s$, the volume of the available space
$V_a(\tau)$ as a function of dimensionless time $\tau$ is recorded
in the very late stages, namely, for the last ten particles added. The saturation
density $\phi_s$ is estimated from this late-stage data by plotting
$\phi(\tau)$ versus $\tau^{-1/d}$ [cf. (\ref{Feder})] and extrapolating
to the infinite-time limit. However, to perform the extrapolation properly
the time increment between each particle addition cannot be taken to be uniform but
instead must increase with increasing time in order to account for the fact
that we only attempt to add particles in the available space. In the very late
stages, this time increment $\Delta \tau$ is given by
\begin{equation}
\Delta \tau = \frac{L^d}{V_a(\tau)}.
\end{equation}
The stopping density $\phi_{stop}$ always
bounds the saturation density
$\phi_{s}$ from below, but as we will soon see, 
$\phi_{stop}$ is very nearly equal to the saturation density $\phi_s$. 

\subsection{Calculation of  the Pair Correlation Function}

We obtain the pair correlation function
$g_{2}(r)$ at the nearly-saturated stopping density $\phi_{stop}$
for a specific configuration by generating a histogram of the average number of 
particle centers $n(r)$ contained in a concentric shell of finite thickness $\Delta r$ at radial distance $r$ 
from an arbitrary reference particle center~\cite{To02a}.  The radial distance $r$ is defined as halfway between 
the inner radius ($r - \Delta r/2$) and the outer radius ($r + \Delta r/2$) of each shell.  The shell thickness 
is termed the bin width.  Let $n_{k}(r)$ represent the accumulated pairs of particles for the entire
system placed in bin $k$ associated with a  radial distance $r$. 
By definition, $n_{k}(r)$ must be an even integer. Then
\begin{equation}
   n(r) = \frac{n_{k}(r)}{N} \hspace{5pt},
\end{equation}
where $N$ is the number of particles in the fundamental cell.  
In general, the pair correlation (or radial distribution) function is defined as 
\begin{equation}
   g_2(r) = \frac{n(r)}{\rho v_{shell}(r)} \hspace{5pt},
\end{equation}
where $v_{shell}$ is the volume of the $d$-dimensional shell, given by
\begin{equation}
   v_{shell} = v_1(r) \left[\frac{(r+\Delta r/2)^d - (r-\Delta r/2)^d}{r^d}\right] \hspace{5pt},
\label{Equation3}
\end{equation}
$\rho$ is the number density $N/L^d$, and $v_1(r)$ is the volume of a $d$-dimensional sphere of radius $r$ as shown 
earlier.  

We compute ensemble-averaged pair correlation functions
by binning up to a maximum distance of
$r_{max}$ for each realization of the ensemble and then averaging over all ensemble members. 
Away from contact, we employ a bin width of $\Delta r = 0.05D$.
Near contact, we use a finer bin width of $\Delta r = 0.005D$
in order to accurately capture the logarithmic divergence of
$g_2(r)$ as the contact value is approached. 

\subsection{Calculation of the Cumulative Coordination Number}

Another quantity of
interest is the cumulative coordination number $Z(r)$,
 which gives the average number of sphere centers within a distance $r$ from a
 given sphere center. It is related to the pair correlation function $g_2(r/D)$ as follows: 
\begin{eqnarray}
Z(r) &=& \rho \int_1^{r/D} s_1(x) g_2(x) dx \nonumber \\
&=& 2^d \, d  \phi \int_{1}^{r/D} x^{d-1} g_2(x) dx,
\label{coord}
\end{eqnarray}
where $s_1(r)=d \pi^{d/2} r^{d-1}/\Gamma(1+d/2)$ 
is the $d$-dimensional surface area of a sphere of radius $r$ \cite{To02a}.  

\subsection{Calculation of the Structure Factor}

Finally, we also compute the structure factor $S({\bf k})$, which provides
a measure of the density fluctuations at a particular wave vector $\bf k$ and is defined
by the relation
\begin{equation}
S({\bf k}) \equiv 1+\rho{\tilde h}({\bf k}),
\label{factor}
\end{equation}
where ${\tilde h}({\bf k})$ is the Fourier transform of the total
correlation function $h({\bf r})\equiv g_2({\bf r})-1$.
When the total correlation in $\mathbb{R}^d$ depends on the wavenumber $k=|\bf k|$, then the structure factor
$S(k)$ in $\mathbb{R}^d$ for any space dimension $d$ is given by \cite{To02c,To06b}
\begin{equation}
S(k)= 1+ \rho\left(2\pi\right)^{\frac{d}{2}}\int_{0}^{\infty}r^{d-1}h(r)
\frac{J_{\left(d/2\right)-1}\!\left(kr\right)}{\left(kr\right)^{\left(d/2\right
)-1}}dr ,
\label{factor2}
\end{equation}
where $J_{\nu}(x)$ is the Bessel function of order $\nu$.

The expression (\ref{factor2}) provides a means for computing the structure
factor by Fourier transforming the real-space total correlation function
in $\mathbb{R}^d$.
If one is interested in the large-wavelength (small $k$) behavior, however, the
large $r$ behavior of $h(r)$ must be known with high precision. Even for relatively large
simulation cells, it is difficult to access
this large-$r$ asymptotic behavior. In such instances, it is better to compute the structure
factor directly from the collective density variables, i.e.,
\begin{equation}
S({\bf k})= \frac{\langle |\rho({\bf k})|^2 \rangle}{N},
\end{equation}
where
\begin{equation}
\rho({\bf k})=\sum_{j=1}^N \exp(i {\bf k}\cdot {\bf r}_j)
\end{equation}
are the collective density variables, angular brackets
denote an ensemble average, and $\bf k$ are the wave vectors appropriate
for the periodic cell of volume $V$.  For the hypercubic fundamental cell
of side length $L$ considered here, the $d$-dimensional wave vectors are given by
\begin{equation}
{\bf k}= \left(\frac{2\pi}{L}n_1,\frac{2\pi}{L}n_2,  \ldots, \frac{2\pi}{L}n_d\right),
\end{equation}
where $n_i$ ($i=1,2,\ldots,d$) are the integers. Thus, the smallest positive
wave vector that one can measure has magnitude $2\pi/L$.
For small to intermediate values of $k$, we will employ the direct method,
while for intermediate to large values of $k$, we will use both the direct
and indirect method [i.e., we calculate $S(k)$ using (\ref{factor2})].

\section{Results and Discussion}

\subsection{Saturation Density}
\label{sat-den}

The saturation density $\phi_s$ for each of the first six space dimensions was determined 
by considering 100-1,000 realizations and several different system
sizes, as discussed in Section III.B. Table \ref{TableI} summarizes 
our results for the saturation density for the largest systems and the associated 
standard error. Included in the table is the stopping density $\phi_{stop}$,
relative system volume $L^d/v_1(1/2)$ for the largest system, where $v_1(1/2)$ is the volume of a hypersphere,
and  the total number of configurations $n_{conf}$.
The results for $d=1,2$ and $3$ agree well
with known results for these dimensions \cite{Re63,Fe80,Hi86,Co88,Ta91}.
We see that the stopping density $\phi_{stop}$ is very nearly equal to
the saturation density $\phi_s$ for all dimensions, except for
$d=1$ where these two quantities are identical. For $d=1$,
no extrapolation was required since we can ensure that
the packings were truly saturated in this instance.

\begin{table}
\begin{center}
\caption{ \label{TableI} The computed saturation density $\phi_{s}$ and
associated standard error for the first six space dimensions.
Included in the table is the stopping density $\phi_{stop}$,
relative system volume $L^d/v_1(1/2)$, and the total number of configurations $n_{conf}$.}
\begin{tabular}{|c|c|c|c|c|}
\hline
Dimension, $d$ & $\phi_{stop}$  &  $\phi_{s}$  & $L^d/v_1(1/2)$ &$n_{conf}$\\ \hline
1   &  0.74750 & 0.74750   $\pm$ 0.000078 & 6688.068486   &  1000 \\
2   & 0.54689  & 0.54700   $\pm$ 0.000063 & 9195.402299   &  1000 \\
3   & 0.38118  & 0.38278   $\pm$ 0.000046 & 13333.333333  &  1000 \\
4   & 0.25318  & 0.25454   $\pm$ 0.000091 & 21390.374000  &  635 \\
5   & 0.16046  & 0.16102   $\pm$ 0.000036 & 66666.666667  &  150 \\
6   & 0.09371  & 0.09394   $\pm$ 0.000048 & 193509.198363 &   75  \\
\hline
\end{tabular}
\end{center}
\end{table}

\begin{figure}[bthp]
\centerline{\psfig{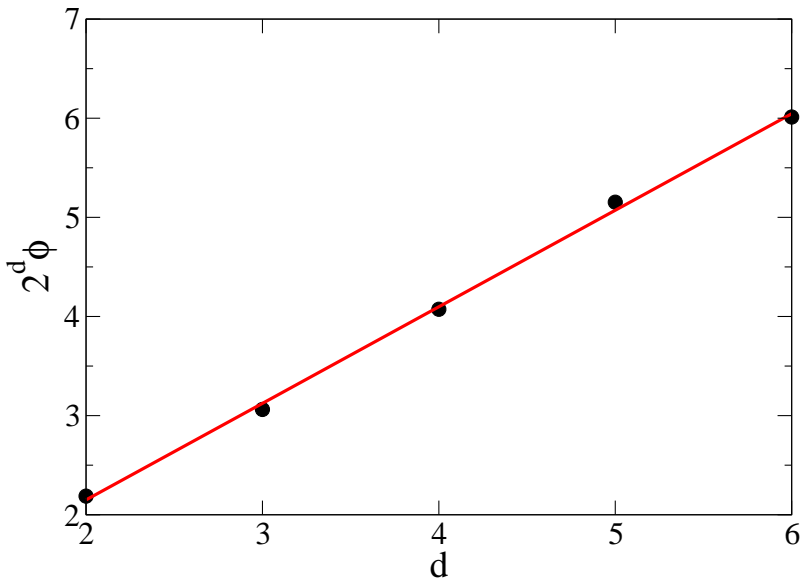}}
\caption{(Color online) Fit of data for the product $2^d \phi_s $ to the linear form  (\ref{linear}) 
for $2 \le d \le 6$. The correlation coefficient is 0.9993, and $c_1=0.202048$ 
and $c_2=0.973872$. }
\label{phi-fit}
\end{figure}

It is of interest to determine how the saturation density $\phi_s$ scales with dimension. 
We already noted that the infinite-time density of
the ghost RSA packing (equal to $2^d$) provides a lower bound on
saturation density of the the standard RSA packing. Therefore, it is natural to consider
the ratio of the saturation density to the  infinite-time density of
the ghost RSA packing, i.e., $2^d \phi_s $. When this ratio is
plotted versus dimension for $2 \le d \le 6$, it is clear that the resulting function,
to an excellent approximation, is linear in $d$, implying
the scaling form
\begin{equation}
\phi_s= \frac{c_1}{2^d}+\frac{c_2 d}{2^d},
\label{linear}
\end{equation}
where $c_1=0.202048$ and $c_2=0.973872$. Indeed, the linear fit of $2^d\phi_s $, shown
in Fig. \ref{phi-fit}, is essentially perfect (the correlation coefficient
is 0.9993). If the data shown in Fig. \ref{phi-fit} is instead fitted up to $d=5$, the
predicted density for $d=6$ from this fit function is within 0.4\% of the corresponding datum.
This indicates that the scaling form for relatively low dimensions is accurate.
In the following subsection, we provide an analytical argument supporting
the same scaling form in the high-dimensional limit. It is noteworthy
that the best rigorous lower bound on the maximal density \cite{Ba92},
derived by considering lattice packings, has the same form as (\ref{linear}).

An interesting conjecture due to Pal{\`a}sti \cite{Pal60}  claims that the saturation
density for RSA packings of congruent, oriented $d$-dimensional cubes  
equals the saturation density 
$\phi_s= 0.747598\ldots$ \cite{Re63} of the one-dimensional problem raised to the $d$th power.
It took over thirty years to show, through a precise Monte Carlo
simulation in two dimensions \cite{Bro91}, that the Pal{\`a}sti conjecture
could not be rigorously true. For RSA packings of congruent, oriented
squares, the saturation density was determined to be $0.562009\pm 0.000004$,
which is close but not equal to $(0.747598\ldots)^2=0.5589\ldots$.
It is noteworthy that the saturation density for oriented
squares is also close to that of circular disks (see Table \ref{TableI}).
These two systems are distinguished from one another in that  
the available space for particle addition in the late stages for the former
are relatively large rectangles,\cite{Bro91} while
for disks they are small ``triangular"-shaped regions \cite{Hi86}.

Is a Pal{\`a}sti-like conjecture (involving raising the one-dimensional
density result to the $d$th power) ever valid for disks? We make the 
simple observation here that the Pal{\`a}sti conjecture is exact
for the ghost RSA packing \cite{To06a} in the infinite-time
limit because $\phi(\infty)=2^{-d}$ in any dimension. Moreover,
it is trivial for us to rigorously prove that a Pal{\`a}sti-like conjecture
cannot be true for the standard RSA packing of spheres in $\mathbb{R}^d$. This
conjecture would state that 
\begin{equation}
\phi_s= \frac{1}{2^{(0.419665\ldots)d}}
\end{equation}
is the saturation density for such a packing for all $d$.
However, this violates the asymptotic Kabatiansky-Levenshtein upper bound (\ref{kab})
for the maximal density of a sphere packing in $\mathbb{R}^d$. 
Therefore, a Pal{\`a}sti-like conjecture cannot be true
for standard RSA packing of spheres $\mathbb{R}^d$. This was known from numerical
experiments, but a proof was never presented until now.

\subsection{Pair Correlation Function}

\begin{figure}
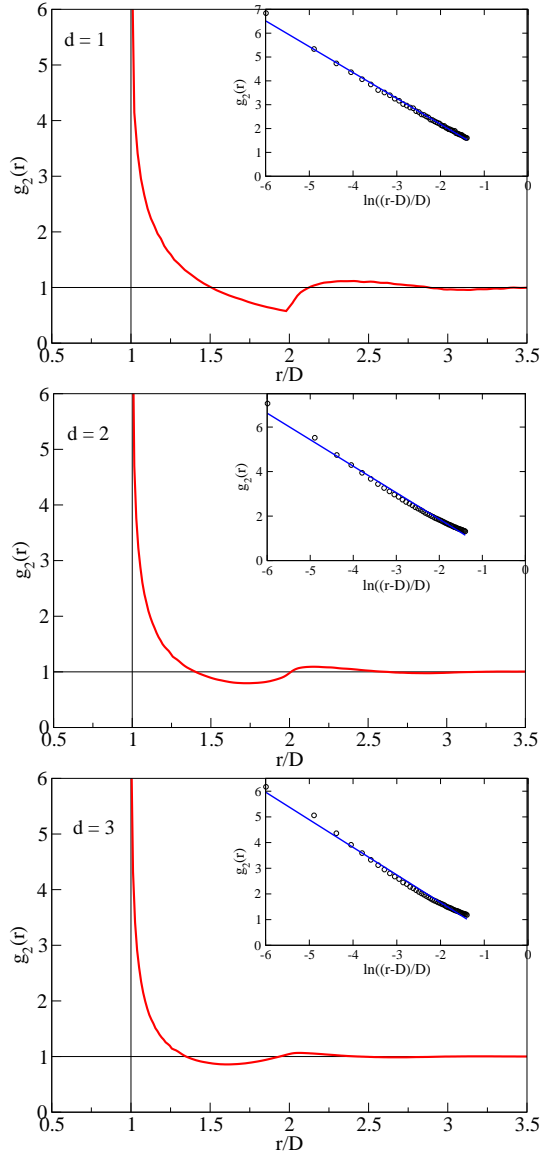

\centerline{\psfig{file=g2r1D.eps,height=2.0in}}
\centerline{\psfig{file=g2r2D.eps,height=2.0in,clip=}}
\centerline{\psfig{file=g2r3D.eps,height=2.0in,clip=}}
\caption{(Color online) The pair correlation functions for RSA packings
for $d=1$ (top panel), $d=2$ (middle panel) and $d=3$ (bottom panel)
very near their respective saturation densities: $\phi=\phi_{stop}=0.74750$, $\phi=
\phi_{stop}=0.54689$
and $\phi=\phi_{stop}=0.38118$, respectively. The insets show semi-logarithmic plots
of the divergence in $g_2(r)$ near contact. The straight
line is a linear fit of the data.
}
\label{g2-1-3}
\end{figure}

\begin{figure}
\centerline{\psfig{file=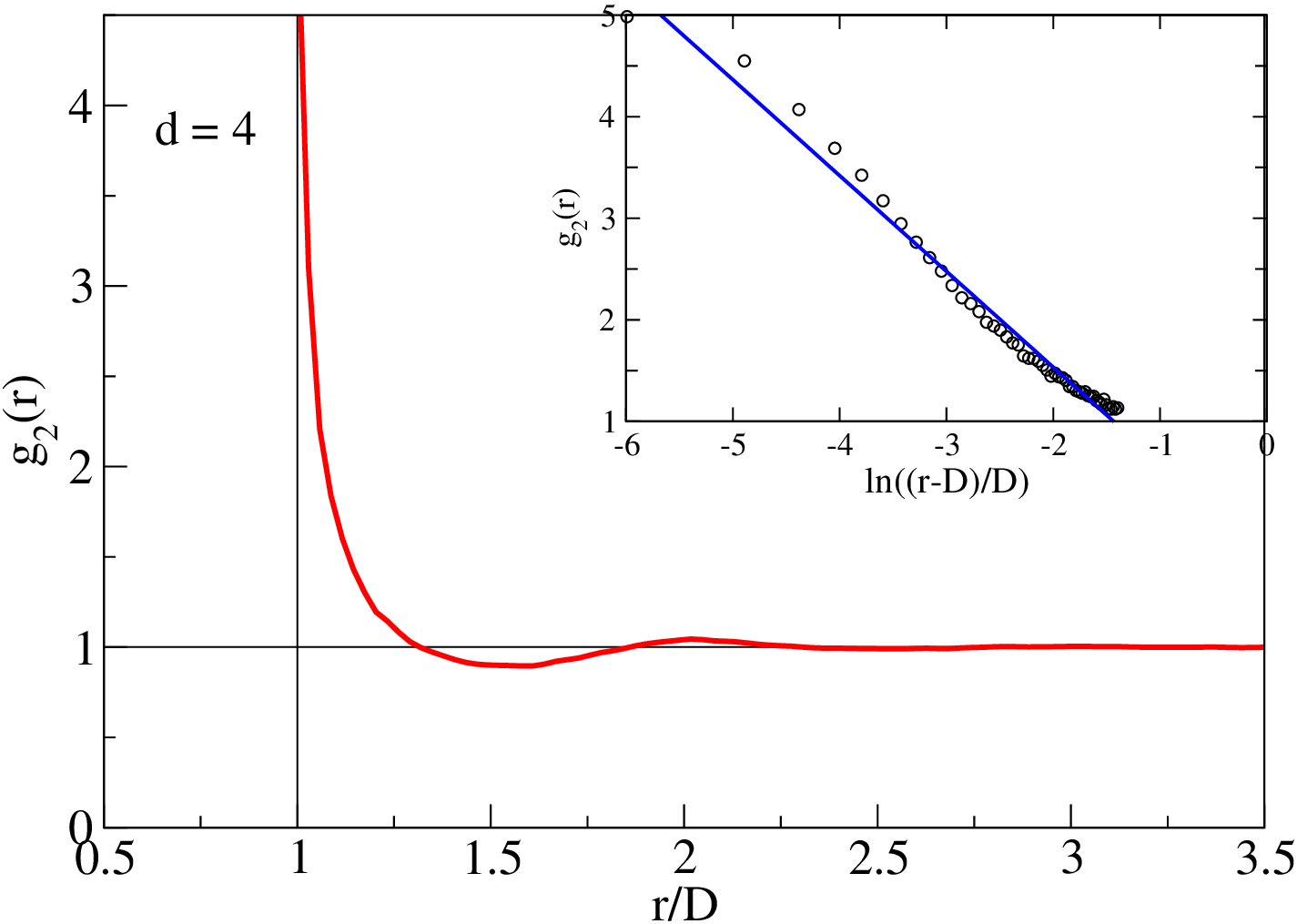,height=2.0in}}
\centerline{\psfig{file=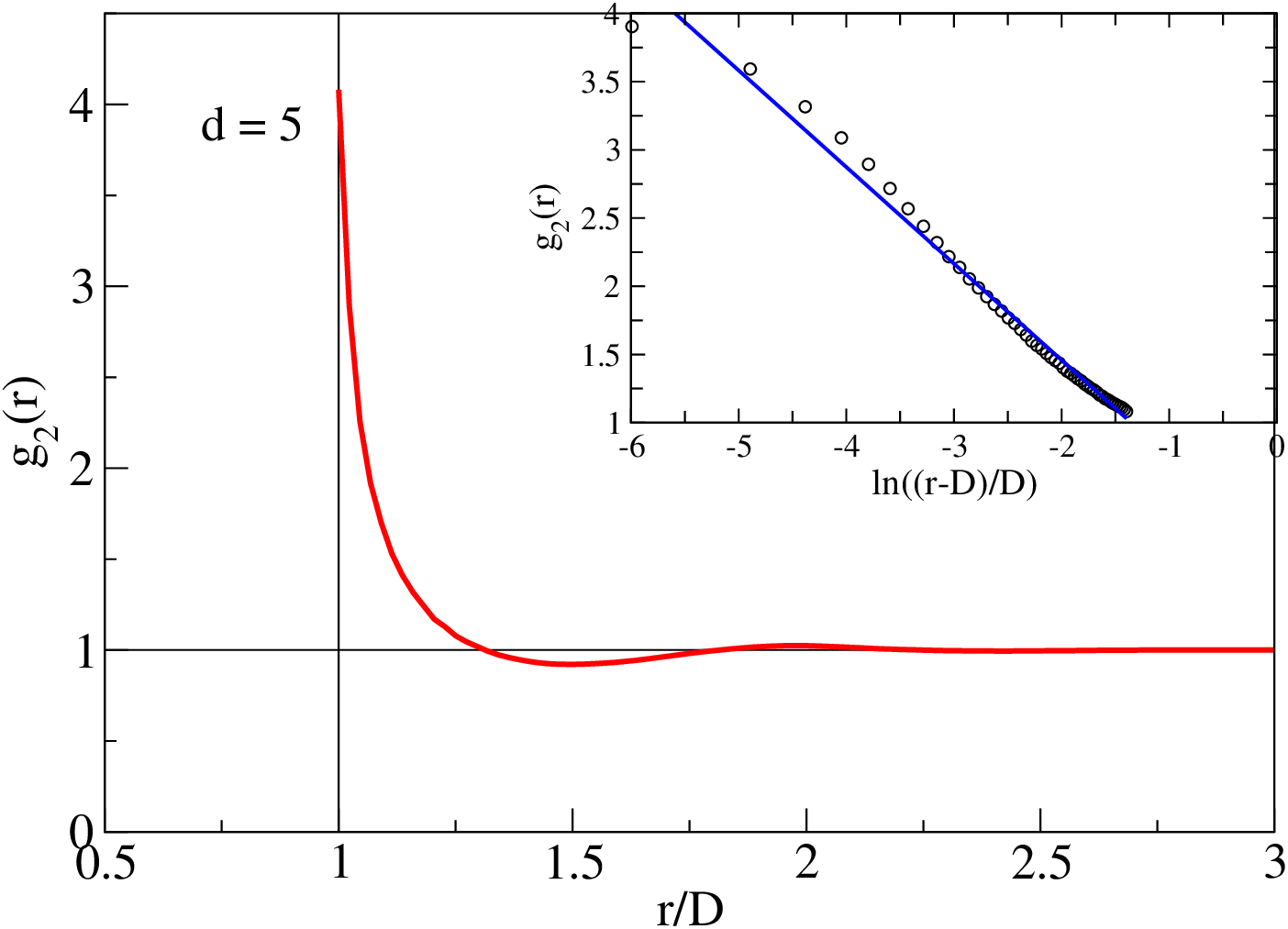,height=2.0in,clip=}}
\centerline{\psfig{file=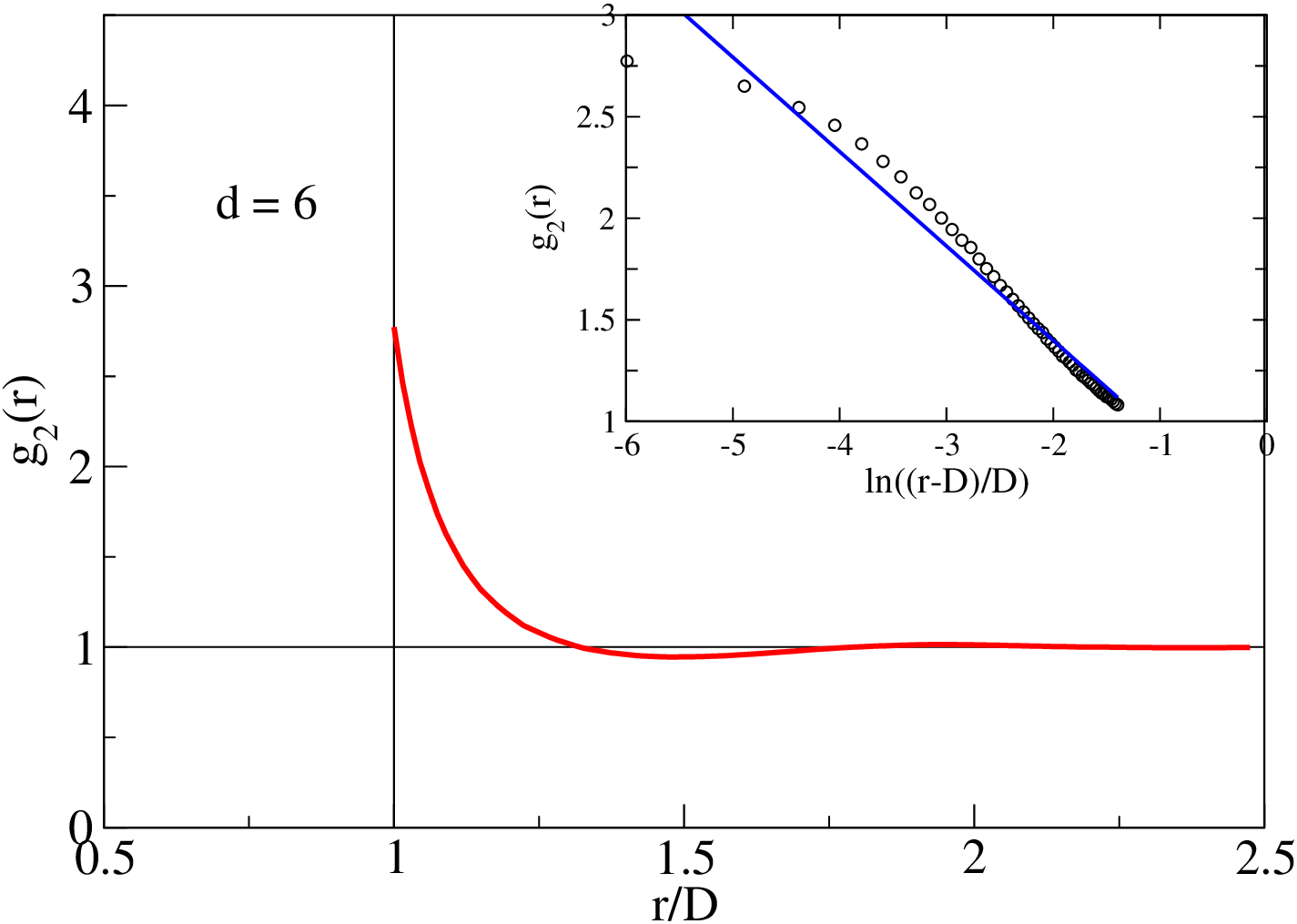,height=2.0in,clip=}}
\caption{(Color online) The pair correlation functions for RSA packings
for $d=4$ (top panel), $d=5$ (middle panel) and $d=6$ (bottom panel) 
very near their respective saturation densities:
$\phi=\phi_{stop}=0.25318$, $\phi=\phi_{stop}=0.16046$
and $\phi=\phi_{stop}=0.09371$, respectively.  The insets show semilogarithmic plots
of the divergence in $g_2(r)$ near contact. The straight
line is a linear fit of the data.
}
\label{g2-4-6}
\end{figure}

Figures \ref{g2-1-3} and \ref{g2-4-6} show the ensemble averaged pair correlation
functions for the first six space dimensions very near 
their respective saturation densities. They are computed from the same
configurations used to calculate the saturation densities, as described
in Section \ref{sat-den}. To our knowledge, our results
for $g_2$ very near the saturation densities have not been
presented before for $d \ge 3$. The inset in each figure shows  the near-contact
behavior, which is consistent with the expected logarithmic divergence
at contact and  fitted to the form:
\begin{equation}
g_2(x)=a_0 \ln(x-1)+a_1, \qquad 1 \leq x \leq 1.135,
\end{equation}
where $x=r/D$. Table \ref{TableII} summarizes the values of the fit parameters
$a_0$ and $a_1$ for each dimension. Of course, the logarithmic
term  overwhelms the constant coefficient $a_1$ 
as $x \rightarrow 1$. Our result for the  logarithmic coefficient
$a_0$ for $d=1$ agrees well with the exact result $a_0=-1.128\ldots$ \cite{Hi86}.
There are no exact results for $a_0$ for $d \ge 2$, but it has
been previously evaluated numerically for $d=2$ by Hinrichsen et al \cite{Hi86},
who obtained the value $a_0=-1.18$, which is somewhat smaller
in magnitude than the value reported in Table \ref{TableII}. These
authors were only able to fit their data in the near-contact region over  about 1.5 decades 
on semi-logarithmic plot due to insufficient statistics. Indeed, we have employed substantially
more configurations than they did and were able to fit our data over about 4.6 decades 
on semi-logarithmic plot. The results reported in Table \ref{TableII}
for $d \ge 3$ have not been presented before. 
Feder et al. also gave an expression for the dominant logarithmic term
for any $d$ in terms of a certain Voronoi statistic and the ``hole-size"
distribution function at contact; but since neither of these quantities are
known analytically, it is not a practically useful relationship.

\begin{figure}[bthp]
\centerline{\psfig{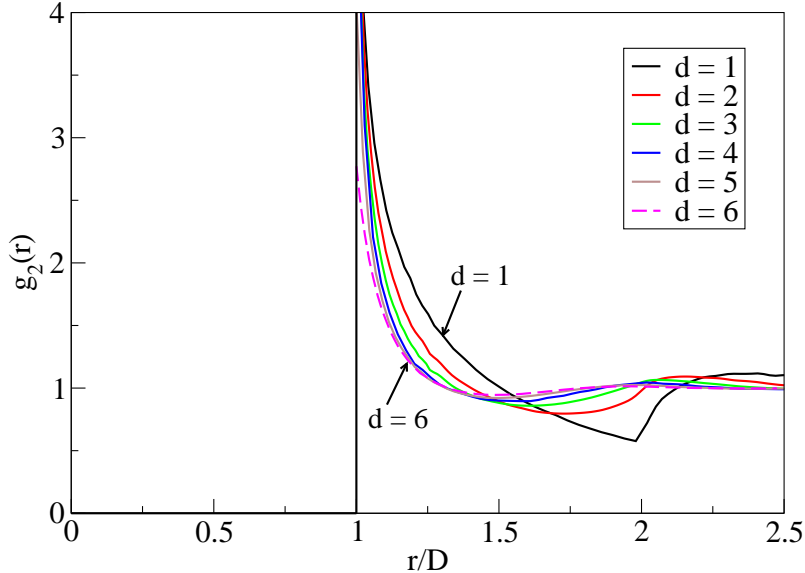}}
\caption{(Color online) The pair correlation functions for RSA packings
for the first six dimensions very near their respective saturation densities. Correlations clearly decrease
as the space dimension increases. Note that the first intercept
of $g_2(r)$ with unity decreases with increasing dimension.}
\label{g2-1-6}
\end{figure}

Figure \ref{g2-1-6} plots all of the pair correlation functions
on the same scale. We see that the ``decorrelation principle,'' which states that unconstrained spatial
correlations diminish as the dimension increases and vanish entirely
in the limit $d\rightarrow \infty$ \cite{To06a,To06b}, is already markedly apparent in these relatively
low dimensions. Correlations away from contact are clearly decreasing
as $d$ increases from $d=1$. The near-contact behavior also is consistent
with the decorrelation principle for $d \ge 2$. Although the logarithmic coefficient
$a_0$ increases in going from $d=1$ to $d=2$, it decreases for all $d >2$ [cf. Table \ref{TableII}].
The decorrelation principle dictates that $a_0$ tends to zero as $d$ tends to infinity.
Similarly, the constant coefficient $a_1$ increases for $d >2$ and for $d=6$ is
equal to $0.661348$ [cf. Table \ref{TableII}]. Indeed, the decorrelation principle requires         
that  $a_1$ tends to unity,
indicating the absence of spatial correlations, as $d$ tends to infinity.

\begin{table}
\begin{center}
\caption{\label{TableII} Fits of the pair correlation function $g_2(r)$ at the very
nearly saturation density $\phi=\phi_{stop}$
to the form $a_0 \ln(x-1) + a_1$ for $1 \leq x \leq 1.135$.}
\begin{tabular}{|c|c|c|}
\hline
Dimension, $d$  &     $a_0$     &    $a_1$    \\ \hline
1               &  -1.11955     &  -0.117475  \\
2               &  -1.29175     &  -0.883021  \\
3               &  -1.16546     &  -0.808843  \\
4               &  -1.00743     &  -0.58044  \\
5               &  -0.714731    &  0.020810  \\
6               &  -0.412100    &   0.661348  \\
\hline
\end{tabular}
\end{center}
\end{table}

\subsection{Cumulative Coordination Number}

The cumulative coordination number $Z(r)$ at $\phi=\phi_{stop}$ 
is easily obtained from the previous results for $g_2(r)$ by performing the integration
indicated in (\ref{coord}). All of these results for $1 \le d \le 6$, to our knowledge,
have not been presented before. For $D \le r \le D(1+\epsilon)$, where $\epsilon \rightarrow 0$,
we can use the asymptotic form to yield the corresponding
expression for $Z(1+\epsilon)$ in any dimension $d$ as
\begin{equation}
Z(1+\epsilon)= 2^d \, d  \phi_s \Big[ a_0\{\ln(\epsilon)-1\}+a_1\Big]\epsilon +{o}(\epsilon),
\label{z}
\end{equation}
where ${o}(\epsilon)$ signifies terms of higher order in $\epsilon$
and distance is measured in units of the hard-sphere
diameter. Thus, $Z(1+\epsilon) \rightarrow
0$ in the limit $\epsilon \rightarrow 0$, i.e., the average contact number
is zero for RSA packings. This feature makes RSA packings distinctly
different from maximally random jammed (MRJ) packings \cite{To00b,Ka02}, 
which have an average contact number equal to $2d$ \cite{Do05d,Sk06}.  This is 
one reason, among others, why the former packing has
a substantially smaller density than the latter. 
The MRJ densities, as determined from computer simulations 
\cite{To00b,Ka02,Do05d,Sk06}, are given  
0.64, 0.46, 0.31 and 0.20 for $d=3,4,5$ and 6, respectively, 
which should  be compared to the RSA saturation densities
given in Table \ref{TableI}. We also note that the appearance
of the product $a_0\ln(\epsilon)\epsilon$ in (\ref{z}) means that the cumulative
coordination number will be concave near contact and possess a positive infinite 
slope at contact.

For values of $r$ away from the near-contact behavior, we can deduce
the following explicit approximation for the $d$-dimensional RSA
cumulative coordination number $Z(x)$ as a function
of the dimensionless distance $x=r/D$ at saturation:
\begin{equation}
Z(x)= (c_1 d+ c_2 d^2)\Bigg[ \Big[a_0\{\ln(\epsilon)-1\}+a_1\Big]\epsilon
+ \frac{x^d}{d}-\frac{(1+\epsilon)^d}{d}\Bigg],
\qquad x \ge 1+\epsilon,
\label{Z-approx}
\end{equation}
where $\epsilon$ is a small positive number (which can be taken to be 0.135 in practice)
describing the range of the near-contact behavior, and  the constants $c_1, c_2, a_0, a_1$
are given in the caption of Fig. \ref{phi-fit} and Table \ref{TableII}.
This approximation is obtained
using the saturation density scaling (\ref{linear}), the near-contact
relation (\ref{z}), and definition (\ref{coord}) employing
the approximation that $g_2(x)=1$, which of course becomes exact
as $x$ and/or $d$ becomes large. In light of the superexponential
decay of $g_2$ in any dimension and the decorrelation principle, $x$ or $d$
does not have to be large for the approximation (\ref{Z-approx}) to be accurate.

\begin{figure}
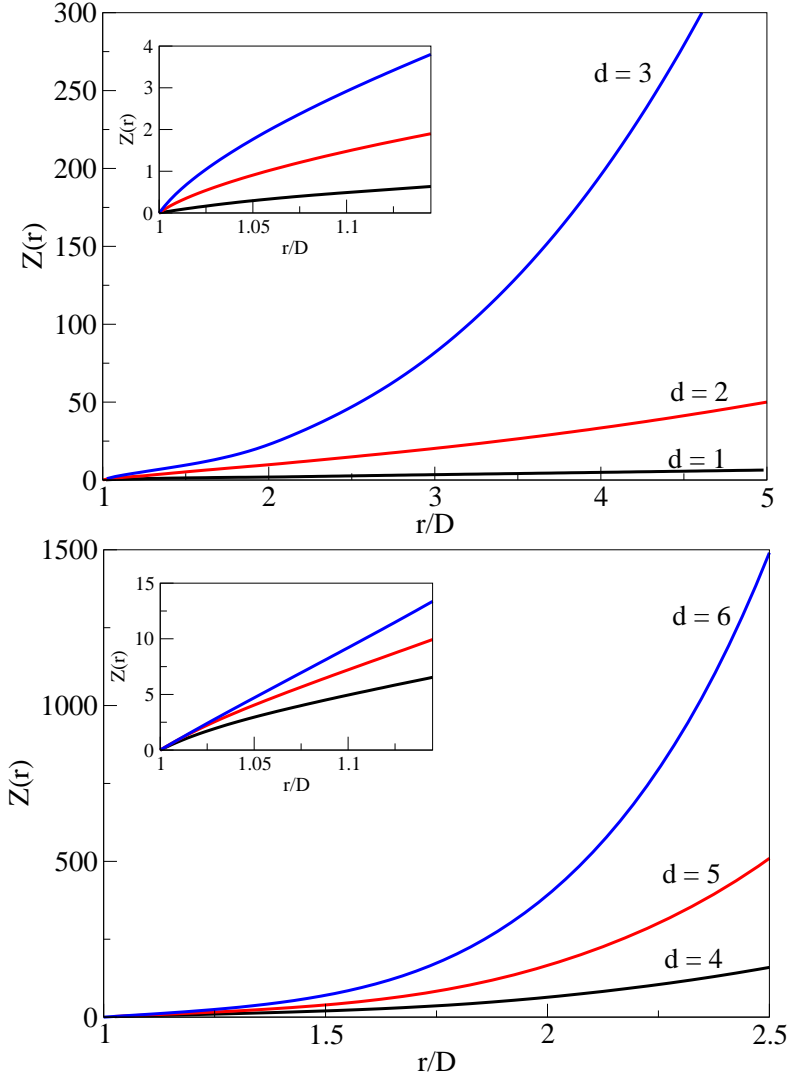

\centerline{\psfig{file=Zr1-3D.eps,height=2.8in}}
\centerline{\psfig{file=Zr4-6D.eps,height=2.8in,clip=}}
\caption{(Color online) The cumulative coordination number $Z(r)$ for RSA packings for
the first six space dimensions very near their respective saturation densities,
as obtained from our numerical data for
$g_2(r)$ and (\ref{coord}). The insets show the concavity of $Z(r)$ near the contact value, which is
in agreement with the  behavior predicted by relation (\ref{z}). Away from contact,
formula (\ref{Z-approx}) provides a good approximation to the numerical
data, especially for $d \ge 3$.}
\label{Z-1-6}
\end{figure}

Figure \ref{Z-1-6} shows the cumulative coordination number $Z(r)$
for the first six space dimensions at their respective saturation densities. These 
results are obtained by numerically integrating (\ref{coord}) using the trapezoidal rule
and our corresponding numerical data for $g_2(r)$.
The insets of these figures clearly show the concavity of $Z(r)$ near contact, as
predicted by (\ref{z}). Away from contact, formula (\ref{Z-approx}) provides a good approximation
to the numerically determined values of $Z(r)$ and is especially accurate for $d \ge 3$.

In Appendix A, we determine kissing (contact) number statistics for saturated RSA 
configurations on the surface of a $d$-dimensional sphere for dimensions $2 \le d \le 5$
and compare to the maximal kissing numbers in these dimensions. It is of interest 
to determine the value of $r$ at which the cumulative coordination number $Z(r)$ 
for a saturated RSA packing in $\mathbb{R}^d$ matches the {\it average} RSA kissing number 
$\langle Z \rangle$ on the surface of a hypersphere in the same dimension. Using Table 
\ref{table4}, we find that $Z(r)=\langle Z \rangle$ for $r/D=1.4233, 1.43042, 1.36044$ 
and $1.3202$ for $d=2,3,4$ and 5, respectively. We see that these distances are relatively
small and decrease with increasing dimension for $2 \le d \le 5$ and would expect the
same trend to continue beyond five dimensions. This behavior is expected because  
RSA packings have superexponential
decay of large-distance pair correlations in any dimension, and the decorrelation principle
dictates that all unconstrained correlations at {\it any} pair distance
must vanish as $d$ becomes large. Therefore, a saturated RSA packing in high dimensions
should be well approximated by an ensemble in which spheres randomly and sequentially
packed in a local region around a centrally  located sphere until
saturation is achieved. Indeed, we have verified this proposition numerically, but do not present such
results here. As the dimension increases, therefore,
the local environment around a typical sphere in an actual
RSA packing in $\mathbb{R}^d$ should be closely
approximated by saturated RSA configurations on the surface
of a $d$-dimensional sphere, even though the former cannot have
contacting particles.

\subsection{Structure Factor}

In light of the discussion given in Section II, the total correlation function $h(r)$ 
decays to zero for large $r$ super-exponentially fast.
It is well known from Fourier transform theory that
if a real-space radial function $f(r)$ in $\mathbb{R}^d$ decreases sufficiently rapidly
to zero for large $r$ such that 
all of its even moments exist, then its Fourier transform ${\tilde f}(k)$ is an even function and analytic
at $k=0$. Thus, the structure factor $S(k)$ for an RSA packing of spheres
in $\mathbb{R}^d$ must be an even, analytic function at $k=0$. Hence,
$S(k)$, defined by  (\ref{factor2}), has an expansion about $k=0$ in any space dimension $d$ for
$0\le \phi \le \phi_s$ of the general form
\begin{equation}
S(k)= S_0 + S_2 k^2 + {\cal O}(k^4),
\label{low-k}
\end{equation}
where $S_0$ and $S_1$  are the $d$-dependent constants defined by
\begin{equation}
S_0=1+ 2^d d \phi \int_0^\infty r^{d-1} h(r) dr \ge 0
\end{equation}
and
\begin{equation}
S_2=-2^{d-1} \phi \int_0^\infty r^{d+1} h(r) dr.
\end{equation}
This analytic behavior of $S(k)$ is to be contrasted with that of sphere packings near the 
MRJ state, which possesses a structure factor 
that is nonanalytic at $k=0$ \cite{Do05d} due to a total
correlation function $h(r)$ having a power-law tail.

It is of interest to determine whether RSA packings are {\it hyperuniform} \cite{To03}
as $\phi \rightarrow \phi_s$ and, if not, their ``distance" from hyperuniformity.
A hyperuniform packing is one in which 
\begin{equation}
\lim_{k\rightarrow 0} S(k)\rightarrow 0,
\end{equation}
i.e., the infinite-wavelength density fluctuations vanish. For RSA packings,
this is equivalent to asking whether the generally nonnegative coefficient $S_0$, defined
in (\ref{low-k}), vanishes. It is known that in one dimension, $S_0 \approx 0.05$ \cite{Bo94},
and hence RSA rods are nearly but not quite hyperuniform. For any non-hyperuniform packing,
the magnitude of $S_0$ provides a measure of its ``distance" from hyperuniformity.
For a Poisson point pattern, it is well known that $S_0=1$, but, in general,
$S_0$ can become unbounded if $h(r)$ decays to zero more slowly
than $r^{-d}$, as it does for a fluid at its critical point.

\begin{figure}
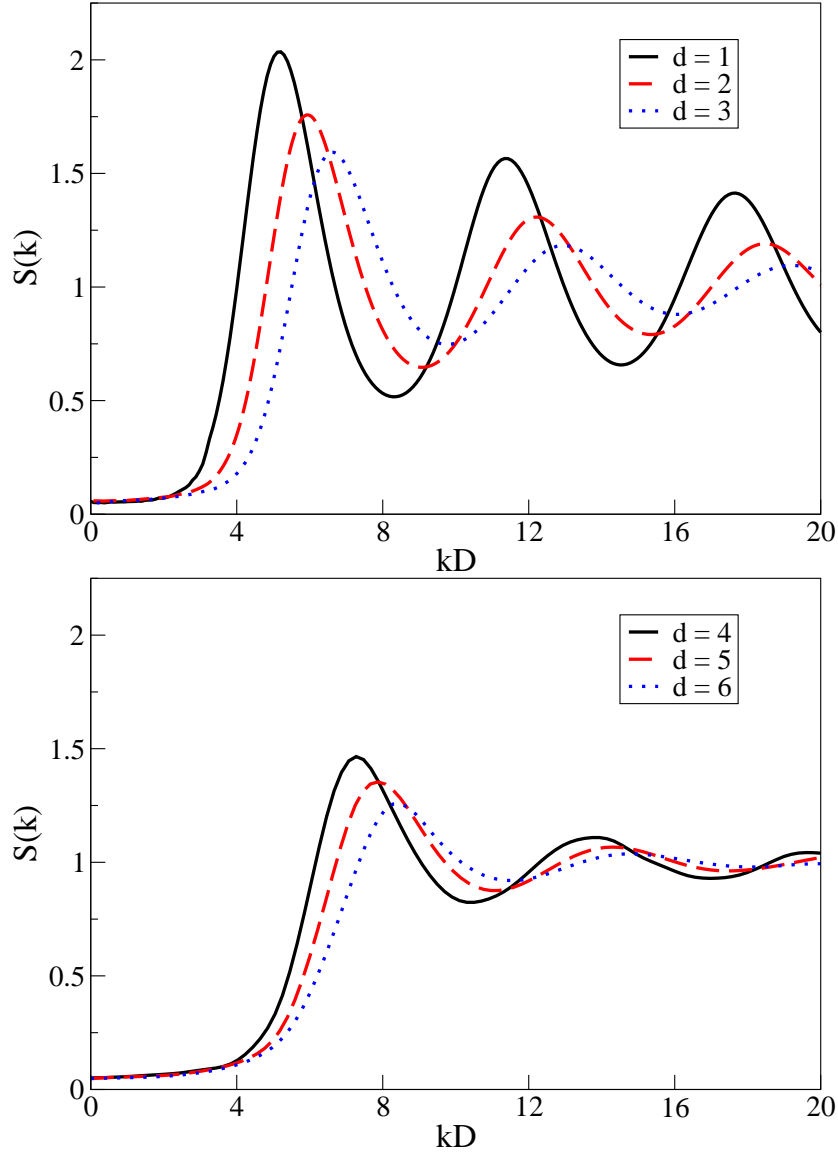

\centerline{\psfig{file=Sofk-1D-3D.eps,height=3.0in,clip=}}
\centerline{\psfig{file=Sofk-4D-6D.eps,height=3.0in,clip=}}
\caption{(Color online) The structure factor $S(k)$ for RSA packings 
for the first six space dimensions very near their respective saturation densities.
Top panel includes curves for $d=1,2$ and $3$ and the bottom
panel includes curves for $d=4,5$ and $6$. Consistent with the behavior
of $g_2$, we see again that pair correlations clearly decrease
as the space dimension increases.}
\label{S-1-6}
\end{figure}

\begin{table}[bthp]
\begin{center}
\caption{\label{S_0} The structure factor $S_0$ at $k=0$ at the stopping density
$\phi_{stop}$ obtained extrapolating our numerical data from the ``direct" method
near $k=0$ using the form (\ref{low-k}) up to quadratic terms.}
\begin{tabular}{|c|c|}
\hline
Dimension, $d$  &     $S_0$       \\ \hline
1               &  0.051      \\
2               &  0.059     \\
3               &  0.050      \\
4               &  0.050      \\
5               &  0.050     \\
6               &  0.050    \\
\hline
\end{tabular}
\end{center}
\end{table}

Our results for $S(k)$ very near the saturation density
for the first six space dimensions are depicted in Fig. \ref{S-1-6}.
To our knowledge, these results for $d \ge 2$ have not been presented before. 
As the space dimension increases, the amplitudes of the oscillations 
of $S(k)$ diminish, consistent with the decorrelation principle.
Note that the minimum value of $S(k)$ in each dimension is achieved at the origin.
Table \ref{S_0} provides the value of the structure factor at $k=0$, denoted by $S_0$,  by extrapolating our numerical data
from the ``direct" method near $k=0$ using the form (\ref{low-k}) up to 
quadratic terms. 
We see that for $1 \le d \le 6$, all of the packings are nearly hyperuniform
and for $d \ge 2$, the distance from hyperuniformity
does not appreciably vary as a function of dimension. In fact, our results
indicate that the minimum value $S_0$ quickly approaches a constant 
value of about $1/20=0.05$ as $d$ becomes large.

The near hyperuniformity of a RSA packing in $\mathbb{R}^d$ at its respective maximal
density is a consequence of the saturation property. Long
wavelength density fluctuations are appreciably suppressed
because spherical gaps of a diameter equal to a sphere diameter
or larger cannot exist in the packing. In an earlier paper \cite{To03}, two
of us (S. T. and F. H. S.) conjectured that saturated but
strictly jammed disordered packings must be hyperuniform, which
was subsequently verified by numerical simulations in three
dimensions \cite{Do05d}. By this reasoning, one would expect a ghost
RSA packing in $\mathbb{R}^d$  not to be hyperuniform, even
at its maximal density of $1/2^d$ because it is never saturated. 
In fact, in Appendix B we determine the structure
factor of the ghost RSA packing exactly and show
that at its maximal density its distance from hyperuniformity increases
as the space dimension $d$ increases, asymptotically approaching
the value of 1/2.

\section{High-Dimensional Scaling of Saturation Density 
and Lower Bounds For Any $d$}

To determine whether the form of the observed density scaling (\ref{linear})
for the first six space dimensions
persists in the high-dimensional limit, we will apply an
optimization procedure that we introduced to
study the structure of disordered packings \cite{To02c,To06b}. We begin
by briefly reviewing this procedure and then apply it to the problem
at hand.  A {\it $g_2$-invariant process} is one in which a given nonnegative pair correlation 
$g_2({\bf r})$ function remains invariant as density varies for all ${\bf r}$ over the range of
densities
\begin{equation}
0 \le \phi \le \phi_*.
\end{equation}
The terminal density $\phi_*$ is the maximum achievable density
for the $g_2$-invariant process subject to satisfaction of
certain necessary conditions on the pair correlation function. 
In particular, we considered those ``test" $g_2(x)$'s that are distributions on $\mathbb{R}^d$ depending
only on the radial distance $x$.
For any test $g_2(x)$, we want to maximize 
the corresponding density $\phi$ satisfying the following three conditions:

\noindent (i) \hspace{0.25in} $g_2(r)  \ge 0 \qquad \mbox{for all}\quad  r,$

\noindent (ii) \hspace{0.2in} $g_2(r)= 0 \qquad \mbox{for}\quad  r<D,$

\noindent (iii)
\begin{displaymath}
\hspace{-0.5in}S(k)= 1+ \rho\left(2\pi\right)^{\frac{d}{2}}\int_{0}^{\infty}r^{d-1}h(r)
\frac{J_{\frac{d}{2}-1}\!\left(kr\right)}{\left(kr\right)^
{\frac{d}{2}-1}}dr \ge 0  \qquad \mbox{for all}\quad  k,
\end{displaymath}
where $S(k)$ is the structure factor defined by (\ref{factor}).
When there exist sphere packings with $g_2$ satisfying conditions
(i)-(iii) for $\phi$ in the interval $[0,\phi_*]$, then we have the lower
bound on the maximal density given by
\begin{equation}
\phi_{\mbox{\scriptsize max}} \ge \phi_*.
\label{true-bound-phi}
\end{equation}

In addition, to the nonnegativity of the
structure factor $S(k)$, there are generally many other conditions that a pair correlation
function of a point process must obey \cite{Cos04}. Therefore, any test $g_2$
that satisfies the conditions (i)--(iii) does not necessarily correspond to
a packing. However, two of us (S. T. and F. H. S.) have
conjectured that a  hard-core nonnegative tempered distribution $g_2({\bf r})$ is a pair correlation function
of a translationally invariant disordered sphere packing \cite{footnote2}  in $\mathbb{R}^d$ 
at number density $\rho$  for sufficiently large $d$  if and only if
$S({\bf k})\ge 0$ \cite{To06a,To06b}. The maximum achievable density
is the terminal density $\phi_*$.

A certain test $g_2$ and this conjecture led to the
putative long-sought exponential improvement on Minkowski's
lower bound \cite{To06a,To06b}.  The validity of this conjecture
is supported by a number of results.
First, the decorrelation principle states that unconstrained
correlations in disordered sphere packings vanish asymptotically in high dimensions
and that the $g_n$ for any $n \ge 3$ can be inferred entirely from a knowledge
of $\rho$ and $g_2$.  Second, the necessary 
Yamada condition \cite{Ya61} appears to only have relevance in very low dimensions.
This states that the variance $\sigma^2(\Omega) \equiv 
\langle (N(\Omega)^2- \langle N(\Omega) \rangle)^2\rangle$
in the number $N(\Omega)$ of particle centers  contained within a region or ``window"
$\Omega \subset \mathbb{R}^d$ must obey the following condition:
\begin{equation}
\sigma^2(\Omega)=\rho |\Omega| \Big[ 1+\rho \int_{\Omega} h({\bf r}) d{\mathbf r}\Big] \ge \theta(1-\theta),
\label{yamada}
\end{equation}
where $\theta$ is the fractional part of the expected number of points 
$\rho |\Omega|$ contained in the window.
Third, we have shown that other new necessary conditions
also seem to be germane only in very low dimensions. 
Fourth, we have recovered the form of known rigorous bounds on the density
in special cases of the test $g_2$  when the
aforementioned conjecture is invoked. Finally, in these latter two instances,
configurations of disordered sphere packings on the flat torus 
have been numerically constructed with such $g_2$ in low dimensions for densities up to the 
terminal density \cite{Cr03,Uc06a}.

We now apply this optimization procedure and the aforementioned conjecture
to ascertain whether the form of the density scaling (\ref{linear})
persists in the high-dimensional limit. In this  limit,
the decorrelation principle as well as our results for the pair
correlation function in the first six space dimensions enable
us to conclude that $g_2(x)$ is very nearly unity for almost
all distances beyond contact except for a very small
nonnegative interval in the near-contact region. Therefore, because
the extra structure in low dimensions representing unconstrained spatial correlations beyond
a single sphere diameter should vanish as $d \rightarrow \infty$, 
we consider a high-dimensional test pair correlation function
in  $\mathbb{R}^d$ that is nonunity within a small positive
interval $1 \le x \le 1+\epsilon$ beyond contact and unity for
all $x$ greater than  $1+\epsilon$, i.e., we consider
\begin{equation}
g_2(x) = \left\{\begin{array}{lll}
0, \qquad & x < 1,\nonumber \\
1+f(x), \qquad  & 1 \le x \le 1+\epsilon, \nonumber \\
1 , \qquad & x>  1,
\end{array}
\right.
\label{step2}
\end{equation}
where $\epsilon$ is a nonnegative constant ($\epsilon \ge 0$) and $f(x)$ is
any integrable function in one dimension that satisfies $f(x)\ge -1$.
This class of functions can include even those that diverge to infinity
as $x \rightarrow 1$. Examples of the latter integrable class include
\begin{equation}
f(x)=-\ln(x-1),
\label{rsa}
\end{equation}
\begin{equation}
f(x)=\frac{1}{(x-1)^\alpha}, \qquad 0 \le \alpha <1,
\label{mrj}
\end{equation}
and
\begin{equation}
f(x)=\delta(x-1),
\label{del}
\end{equation}
where $\delta(x)$ is the Dirac delta function.
Equation (\ref{rsa}) describes the divergence seen in $g_2(x)$ of the standard 
RSA packing at contact [cf. (\ref{rsa-contact})].
By contrast, Eq. (\ref{mrj}) characterizes the near-contact divergence
of $g_2(x)$ for maximally random jammed (MRJ) packings \cite{To00b}
with $\alpha \approx 0.6$ \cite{Do05c,Sk06}. Equation (\ref{del}) describes
random sphere packings with a positive average contact number.    
For general $f(x)$, the corresponding structure factor [cf. (iii)] for the test function
(\ref{step2}) in any dimension $d$ is given by 
\begin{equation}
S(k)=1 -\frac{ 2^{\frac{3d}{2}} \,\phi \Gamma(1+d/2) }{k^{\frac{d}{2}}}\left[ J_{\frac{d}{2}}(k)-k \int_1^{1+\epsilon}
x^{\frac{d}{2}} f(x)
J_{\frac{d}{2}-1}(kx)dx\right],
\label{rsa-factor}
\end{equation}
where $\nu=d/2$.

We now make use of the following
general result that applies to any function $G(x)$ that is bounded
as $x \rightarrow 1$:
\begin{equation}
\lim_{\epsilon\rightarrow 0} \int_{1}^{1+\epsilon} G(x) f(x) dx = G(1)I(\epsilon)
\label{limit}
\end{equation}
where $I(x)$ is the indefinite integral
\begin{equation}
I(x)= \int f(x) dx.
\end{equation}
For the functions (\ref{rsa}), (\ref{mrj}) and (\ref{del}), the integral $I(x=\epsilon)$ is
respectively given by
\begin{equation}
I(\epsilon)=(1-\ln \epsilon)\epsilon,
\end{equation}
\begin{equation}
I(\epsilon)=\frac{\epsilon^{1-\alpha}}{1-\alpha}, \qquad 0 \le \alpha <1.
\end{equation}
and 
\begin{equation}
I(x)=1.
\end{equation}

Making use of the result (\ref{limit}) in (\ref{rsa-factor}) 
yields the structure factor to be given by
\begin{equation}
S(k)=1 -\frac{ 2^{\frac{3d}{2}}\phi \, \Gamma(1+d/2) }
{k^{\frac{d}{2}}}\left[ J_{\frac{d}{2}}(k)-k J_{\frac{d}{2}-1}(k)I(\epsilon)\right].
\label{high-d-S}
\end{equation}
The structure factor for small $k$ can be expanded in a MacLaurin series
as follows:
\begin{equation}
S(k)=1+2^d\phi[d I(\epsilon)-1]+ \frac{2^{d-1}\phi}{d+2}[1-I(\epsilon)(d+2)]\,k^2
+{\cal O}(k^4).
\end{equation}
The last term changes sign if $I(\epsilon)$ increases past $1/(d+2)$.
At this crossover point,
\begin{equation}
S(k)=1-\frac{2^{d+1}}{d+2}\phi+ {\cal O}(k^4)
\label{factor3}
\end{equation}
Under the constraint that the minimum of $S(k)$ occurs at $k=0$, the terminal
density is then given by
\begin{equation}
\phi_*=\frac{d+2}{2^{d+1}}(1-S_0),
\label{term2}
\end{equation}
where $S_0\in[0,1]$ is the value of the structure factor at $k=0$ or the assumed
minimum value in the high-dimensional limit.
Thus, we see that the terminal density is independent of the specific
form of $I(\epsilon)$ or, equivalently, the choice of the function $f(r)$ [cf. (\ref{step2})],
which only has influence in a very small nonnegative interval around contact.
For a  hyperuniform situation ($S_0=0$), the formula (\ref{factor3})  reduces to 
\begin{equation}
\phi_*=\frac{d+2}{2^{d+1}},
\label{term3}
\end{equation}
which was obtained previously \cite{To02c} for the specific choice of $f(r)$ given 
by (\ref{del}). 

It is noteworthy that the high-dimensional asymptotic structure factor 
relation (\ref{high-d-S}) under the conditions leading to (\ref{term2}) yields a 
structure factor for $d=6$, a relatively low dimension, 
that is remarkably close to our corresponding simulational
RSA result (depicted in the bottom panel of Fig. \ref{S-1-6}) for most
values of the wavenumber $k$.
Such agreement between the asymptotic and
low-dimensional numerical results strongly suggests that our
asymptotic form (\ref{step2}) for the pair correlation function  
indeed captures the true high-dimensional behavior for RSA packings.

\begin{figure}[bthp]
\centerline{\psfig{file=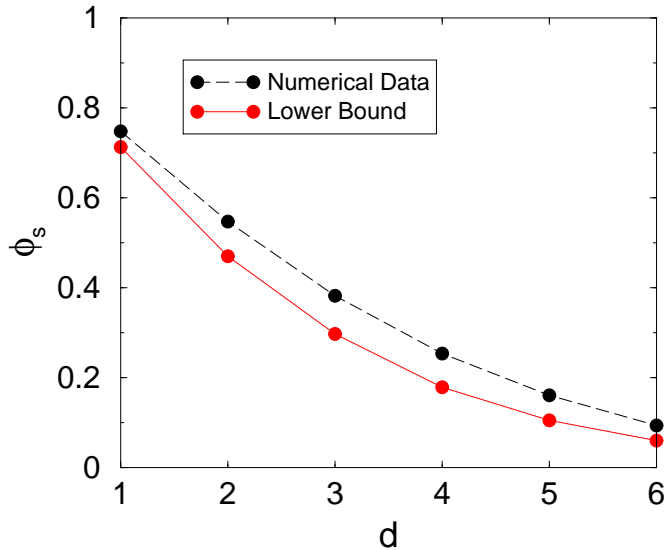,height=3.0in,clip=}}
\caption{(Color online) Comparison of the lower bound (\ref{rsa-bound}) on
the saturation density (with $S_0=0.05$) to our corresponding numerical
data for the first six space dimensions.}
\label{lower-bound}
\end{figure}

In summary, we see that the high-dimensional
result (\ref{term2}) shows that the form of the density scaling (\ref{linear})
at saturation for relatively low-dimensional RSA packings persists in the high-dimensional limit. 
Indeed, as we will show below, the high-dimensional scaling (\ref{term2}) provides a lower bound
on the RSA saturation density $\phi_s$ for arbitrary $d$, i.e.,
\begin{equation}
\phi_s \ge \frac{d+2}{2^{d+1}}(1-S_0).
\label{rsa-bound}
\end{equation}
Figure \ref{lower-bound} depicts a graphical
comparison of the lower bound (\ref{rsa-bound}) to our numerical 
data for the saturation density for the first six space dimensions.

The lower bound (\ref{rsa-bound}) is a consequence of a 
more general principle that enables us to exploit
high-dimensional information in order to infer scaling behavior
in low dimensions, as we now describe.
For any particular packing construction in $\mathbb{R}^d$
(e.g., RSA, ghost RSA or MRJ packings), the highest 
achievable density $\phi_m(d)$ decreases with increasing dimension $d$.
Therefore, the scaling for the maximal density in the asymptotic limit
$d \rightarrow \infty$, which we denote by $\phi_m(\infty)$, provides a lower bound on $\phi_m(d)$ for
any finite dimension, i.e.,
\begin{equation}
\phi_m(d) \ge \phi_m(\infty).
\label{gen-bound}
\end{equation}

In the case of RSA packings in $\mathbb{R}^d$,
we have shown that the high-dimensional density scaling is provided
by the analysis leading to (\ref{term2}) and therefore use of (\ref{gen-bound}) 
yields the lower bound (\ref{rsa-bound}) in which $S_0$ is a small positive
number. It is noteworthy that a lower bound on the MRJ density $\phi_{MRJ}$
is given by the right side of inequality of (\ref{rsa-bound}) but with $S_0=0$,
i.e., 
\begin{equation}
\phi_{MRJ} \ge \frac{d+2}{2^{d+1}}.
\label{mrj-bound}
\end{equation}
This is obtained by recognizing that the same high-dimensional scaling
analysis as we used for RSA applies with one qualitative difference. The high-dimensional
limit of the pair correlation function of an MRJ packing is expected to be
of the same form as (\ref{step2}) but where $f(x)$ is a Dirac delta
function to account for interparticle contacts due to the constraint
of jamming. Since we know that MRJ packings are hyperuniform ($S_0=0$) \cite{Do05d,Sk06},
then (\ref{term2}) together with (\ref{gen-bound}) produces the  bound
(\ref{mrj-bound}). The MRJ lower bound (\ref{mrj-bound}) yields
0.3125, 0.1875, 0.109375, 0.0625 for $d=3,4,5$ and 6, respectively,
which is to be compared to the corresponding actual
MRJ densities of 0.64, 0.46, 0.31 and 0.20
\cite{To00b,Ka02,Do05d,Sk06}. Note that in Ref. \cite{To06b},
the right side of (\ref{mrj-bound}) was argued to be a lower bound
on the maximal density $\phi_{\mbox{\scriptsize max}} $ of {\it any} sphere packing
in $\mathbb{R}^d$. In the case,
of the ghost RSA packing, we know that $1/2^d$ is the maximal density $\phi_{GRSA}$
for any dimension, and therefore this result in conjunction
with (\ref{gen-bound}) yields the lower bound $\phi_{GRSA} \ge 1/2^d$,
which of course is exact.

\section{Conclusions}

We have studied the structural characteristics  of random sequential
addition (RSA)  of congruent spheres in $d$-dimensional Euclidean space 
$\mathbb{R}^d$ in the infinite-time or saturation limit for the first six
space dimensions ($1 \le d \le 6$) both numerically
and theoretically. Specifically, by numerically generating saturated RSA
configurations in each of these dimensions, we determined the saturation 
density, pair correlation function, cumulative coordination number and the structure factor.
We found that for $2 \le d  \le 6$, the saturation density $\phi_s$ 
has the scaling given by (\ref{linear}).
Using theoretical considerations, we showed analytically that the same density-scaling form persists 
in the high-dimensional limit. 
A byproduct of the aforementioned high-dimensional analysis was the 
determination of a relatively sharp lower bound on the saturation density (\ref{rsa-bound})
of RSA packings  for any $d$,
which utilized the infinite-wavelength limit of the structure factor
in the high-dimensional limit. Thus, high-dimensional
information was exploited to provide density estimates
in low dimensions. We proved rigorously that a Pal{\`a}sti-like conjecture
cannot be true for RSA hyperspheres. We also demonstrated that the structure
factor $S(k)$  must be analytic at $k=0$ and that RSA packings for $1 \le d \le 6$
are nearly ``hyperuniform" (i.e., infinite wavelength density fluctuations
vanish). Consistent with the recent ``decorrelation principle," we find
that pair correlations markedly diminish as the space dimension increases up to six.

In Appendix A, we obtained kissing number statistics for saturated RSA configurations on the surface
of a $d$-dimensional sphere for dimensions $2 \le d \le 5$
and compared to the maximal kissing numbers in these dimensions. The discrepancy between
average RSA kissing numbers and maximal kissing number was found to increase as the space dimension
increased. Finally, in Appendix B, we determined the structure factor exactly for the related ``ghost"
RSA packing in $\mathbb{R}^d$ and showed that its distance from ``hyperuniformity"
increases as the space dimension increases, approaching a constant
asymptotic value of $1/2$. 

It is interesting to observe 
that the best known rigorous lower bound on the maximal density \cite{Ba92},
derived by considering Bravais lattice packings, has the same form as the density scaling
(\ref{linear}) for RSA packings, i.e.,
for large $d$, it is dominated by the term $d/2^d$. The fact that the
saturation density of disordered RSA packings approaches this rigorous lower bound
suggests the existence disordered packings whose density
surpasses the densest lattice packings in some sufficiently high dimension.
The reason for this is that we know that there are disordered packings
in low dimensions whose density exceeds that of corresponding saturated RSA packings
in these dimensions, such as maximally random jammed (MRJ)
packings \cite{To00b,Ka02,Do05d,Sk06}. The density of saturated RSA packings in dimension $d$ is substantially smaller than
the corresponding MRJ value because, unlike the latter packing,
the particles can neither  rearrange nor jam. The possibility that disordered
packings in sufficiently high dimensions are the densest is consistent with
a recent conjectural lower bound on the density of disordered hard-sphere
packings that was employed to provide the putative
exponential improvement on Minkowski's 100-year-old bound \cite{To06b}. The asymptotic behavior 
of the conjectural lower bound is controlled by $2^{-(0.77865\ldots) d}$.

Challenging problems worth pursuing in future work
are the determinations of analytical constructions of disordered sphere
packings with densities that equal or exceed $d/2^d$ for sufficiently large $d$
or, better yet, provide exponential improvement on Minkowski's lower bound.
The latter possibility would add to the growing evidence that disordered packings
at and beyond some sufficiently large critical dimension might be
the densest among all packings. This scenario would imply  
the counterintuitive existence of disordered classical ground states for some continuous potentials
in such dimensions. 

\begin{acknowledgments}

We thank Henry Cohn for helpful comments
about maximal kissing number configurations.
This work was supported by the Division of Mathematical Sciences at
NSF under Grant No. DMS-0312067. O. U. U. gratefully acknowledges the support 
of the Department of Energy CSGF program.

\end{acknowledgments}

\begin{appendix}

\section{RSA Kissing Number}

The number of unit spheres in $\mathbb{R}^d$ that simultaneously touch 
another unit sphere without overlap is called the {\it kissing number} (also
known as the contact number or coordination number). 
The {\it kissing number problem} seeks the maximal kissing
number $Z_{\mbox{\scriptsize max}}$ as a function of $d$.
In dimensions $1 \le d \le 3$, the maximal kissing numbers
are known and correspond to the kissing numbers of the densest
sphere packings, which are Bravais lattices \cite{Co93,Ha05}:
the simple linear lattice for $d=1$, the triangular lattice
for $d=2$, and the face-centered cubic (FCC) lattice for $d=3$.
For $d=1$ and $d=2$, the maximal kissing numbers are exactly 2 and 6, respectively.
Although one dimension is trivial, it is a ``miracle" of two
dimensions that 6 circles can simultaneously touch another circle
without any gaps, and in this sense this unique configuration (up to trivial
rotations) is ``rigid" because there are no displacements
of the 6 contacting circles that lead to a different configuration
while maintaining the contacts.
This unique kissing number arrangement is also six-fold symmetric. 
In three dimensions, it is known that 
$Z_{\mbox{\scriptsize max}}=12$, which is achieved by the FCC sphere
packing, but there are no unique configurations
because gaps exist between contacting spheres that enable one optimal 
kissing configuration to be displaced
into a different optimal configuration, and therefore optimal configurations
need not have any symmetry. The aforementioned subtleties in the three-dimensional case
was at the heart of a famous debate in 1694 between Issac Newton 
(who claimed that $Z_{\mbox{\scriptsize max}}=12$)
and David Gregory (who contended that $Z_{\mbox{\scriptsize max}}=13$).

One of the generalizations of the FCC lattice to higher dimensions is the $D_d$
checkerboard lattice, defined by taking a cubic lattice and placing
spheres on every site at which the sum of the lattice indices is even
{\em i.e.,} every other site.  The densest packing for $d=4$ is
conjectured to be the $D_4$ lattice, with a kissing number $Z = Z_{\mbox{\scriptsize max}} =
24$~\cite{Co93}, which is also the maximal kissing number in
$d=4$~\cite{Mu04}.  This optimal configuration
is referred to as the 24-cell, which is both rigid and highly symmetric.
For $d=5$, the densest packing is conjectured to
be the $D_5$ lattice with kissing number $Z =
40$~\cite{Co93}. This kissing configuration is also highly symmetrical.
The maximal kissing numbers $Z_{\mbox{\scriptsize max}}$ for $d=5$
is not known, but has the following bounds: $40 \leq
Z_{\mbox{\scriptsize max}} \leq 46$.

Here we determine the distribution of kissing numbers $Z_i$
by placing hyperspheres randomly and sequentially on the surface 
of a hypersphere at the origin until the surface is saturated.
We call such a configuration a saturated RSA kissing number configuration.
The average kissing number $\langle Z \rangle$ is given by
\begin{equation}
\langle Z \rangle = \sum_{i=1} Z_i P(Z_i),
\end{equation}
where $P(Z_i)$ the probability of finding a saturation kissing number $Z_i$.
We begin our simulations by placing a central hypersphere of {\it unit diameter}
at the origin of a hypercubic simulation box.  A large number of points $n_{pts}$ ($n_{pts}= 2 \times 10^5$ 
for $d=2$, $n_{pts}=10^6$ for $2 \le d \le 4$,  and $10^7$ points for $d=5$) 
are uniformly distributed in the region between the  exclusion hypersphere of {\it unit radius} surrounding
the hypersphere of unit diameter and the boundary of the simulation box. 
Each point is randomly and sequentially radially projected (in the direction
of the hypersphere center) to the surface of the exclusion hypersphere
(via a radial distance rescaling) subject to the nonoverlap condition, i.e., a  projected point is accepted 
if the angular separation between it and any other previously accepted point
is greater than or equal to 60 degrees, otherwise it is rejected. The simulation terminates when all projected 
points obey this nonoverlap condition at which time the exclusion-sphere
surface is taken to be saturated, i.e., the surface of the central sphere of unit
diameter is saturated with contacting spheres of unit diameter.
We found that the number of points $n_{pts}$  that we initially distributed in the simulation
box before the projection step is sufficiently large to ensure that the
surface of the hypersphere is truly saturated after the projection step in the dimensions considered.

\begin{table}
\begin{center}
\caption{\label{table4}Kissing number statistics for saturated RSA 
configurations on the surface of a $d$-dimensional sphere for dimensions 
$2 \le d \le 5$. Here $Z_i$ is the integer-valued saturation kissing number, 
$P(Z_i)$ the probability of finding a saturation kissing number $Z_i$, 
$\langle Z \rangle$ the average saturation kissing number. In each dimension, the statistics
are determined from 100,000 configurations. We also
include the largest known kissing numbers $Z_{\mbox{\scriptsize max}}$.}
\scriptsize
\begin{tabular}{|c|cc|c|c|}
\hline
Dimension, $d$ & Kissing Number, $Z_i~$   & ~Probability, $P(Z_i)$  & 
$\langle Z \rangle$  &
$Z_{\mbox{\scriptsize max}}$ \\ \hline
   2   &     4             &   0.515680    &   4.48432   &   6     \\
       &     5             &   0.484320    &             &         \\
\hline
    3 &     6             &   0.001400    &   8.34957   &   12    \\
      &     7             &   0.091020    &             &         \\
      &     8             &   0.502230    &             &         \\
      &     9             &   0.367400    &             &         \\
      &     10            &   0.037860    &             &         \\
      &     11            &   0.000090    &             &         \\
\hline
   4  &     11            &   0.001840    &   13.80530  &   24    \\
      &     12            &   0.055960    &             &         \\
      &     13            &   0.302440    &             &         \\
      &     14            &   0.435910    &             &         \\
      &     15            &   0.183060    &             &         \\
      &     16            &   0.020260    &             &         \\
      &     17            &   0.000520    &             &         \\
      &     18            &   0.000010    &             &         \\
\hline
   5  &     17            &   0.000030    &   21.46765  &   40    \\
      &     18            &   0.001530    &             &         \\
      &     19            &   0.024510    &             &         \\
      &     20            &   0.146930    &             &         \\
      &     21            &   0.341250    &             &         \\
      &     22            &   0.329250    &             &         \\
      &     23            &   0.132350    &             &         \\
      &     24            &   0.022290    &             &         \\
      &     25            &   0.001810    &             &         \\
      &     26            &   0.000050    &             &         \\
\hline
\end{tabular}
\end{center}
\end{table}

Table \ref{table4} provides kissing number statistics for saturated RSA configurations
for dimensions $2 \le d \le 5$. For $d=2$, only two values of $Z_i$ are allowed:
4 and 5. Any kissing number equal to 3 or less is prohibited because
such configurations are not saturated. On the other hand, a kissing number
of 6 has a probability of zero of occurring by random sequential addition because
the configuration corresponding to this optimal kissing number is unique.
We find that the average kissing number is approximately equal to 4.5. For $d=3$,
the average kissing number $\langle Z \rangle=8.34135$. The fact
that the maximal kissing number configurations ($Z_{\mbox{\scriptsize max}}=12$)
in $\mathbb{R}^3$ are nonunique implies that a configuration
of 12 spheres has a positive (albeit small) probability of
occurring via an RSA process. Nonetheless, we were not able
to observe such a configuration in a total of $10^6$ configurations.
The smallest observed kissing number for $d=3$ was six, which
presumably is the smallest number required for a saturated packing.
The minimal kissing number configurations for saturation are related to the following problem: How can $n$ points 
be distributed on a unit sphere such that they maximize the minimum distance between any pair of points?
For six points, the solution to his problem is well known: they should be placed at the  
vertices of an inscribed regular octahedron. Since the minimum angular separation between
any pair of points in this highly symmetric case is 90 degrees, the associated kissing number configuration
is saturated. Note that highest kissing number of 18 reported for $d=4$ is substantially
smaller than the maximal kissing number $Z_{\mbox{\scriptsize max}}=24$.
Apparently, achieving kissing numbers that approach those of the  optimal highly 
symmetric, rigid 24-cell configuration by a random sequential addition is effectively impossible.
Therefore, it is plausible that the possible configurations corresponding
to kissing numbers of 19-23 are also characterized by
high degree of symmetry (and possibly rigidity) based upon the absence
of such kissing numbers. The fact that the average
kissing number for $d=5$ is substantially lower than the
highest known kissing number of 40 is presumably related
to the high symmetries required to achieve high $Z$ values
in this dimension.

\begin{figure}
\centerline{\psfig{file=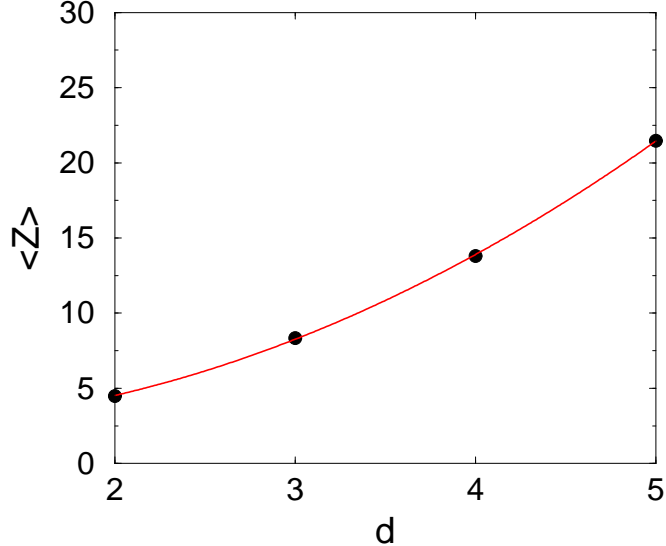,height=3.0in,clip=}}
\caption{(Color online) Our numerical data (black circles) for the average kissing number
$\langle Z \rangle$ as a function of dimension $d$ and the quadratic 
fit function (\ref{fit}) (solid curve).}\label{kiss}
\end{figure}

Our data for the average RSA kissing number 
over the range of considered dimensions is fit very well
by the following quadratic expression in $d$:
\begin{equation}
\langle Z \rangle = b_0+b_1\,d + b_2 \,d^{\,2},
\label{fit}
\end{equation}
where $b_0=2.74488$, $b_1=-1.01354$ and $b_2=0.950395$, and the
correlation coefficient is 0.9999. The data and this quadratic fit function 
are depicted in Figure \ref{kiss}. If this expression
persisted for large $d$, it would predict that the average
RSA kissing number asymptotically grows as $d^2$.

\section{Structure Factor for Ghost RSA Packings}

\begin{figure}
\centerline{\psfig{file=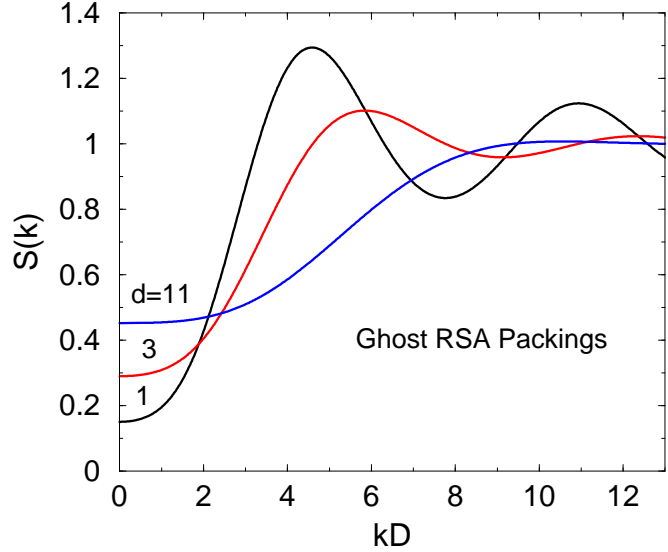,height=3in}}
\caption{(Color online) The structure factor $S(k)$ versus $kD$ for
various space dimensions ($d=1,3$ and $11$)
for ghost RSA packings.}
\label{ghosts}
\end{figure}

For ghost RSA packings of spheres
of diameter $D$ in $\mathbb{R}^d$, the pair correlation function in
the infinite-time limit is given exactly for any space dimension $d$ by the following expression \cite{To06a}:
\begin{equation}
g_2(r)=\frac{\Theta(r-D)}{1-\alpha_2(r;D)/2},
\label{g2-3}
\end{equation}
where $\Theta(x)$ is the unit step function, 
equal to zero for $x<0$ and unity for $x \ge1$,
and $\alpha_2(r;D)$ is the intersection volume of two spheres
of radius $D$ whose centers are separated by the distance $r$
divided by the volume of a sphere of radius $D$. Expressions for the
{\it scaled} intersection volume $\alpha_2(r;D)$ for any $d$ are known
exactly; see Refs.  \cite{To06b} and \cite{To03} for two different representations. 
The scaled intersection volume $\alpha_2(r;D)$ takes its maximum value
of unity at $r=0$ and monotonically decreases with increasing $r$
such that it is nonzero for $0 \le r <2D$, i.e., it has compact
support. The corresponding total correlation $h(r)=g_2(r)-1$ is
given by
\begin{equation}
h(r)=-\Theta(D-r)+\frac{\alpha_2(r;D)}{2-\alpha_2(r;D)}\Theta(r-D)\Theta(2D-r).
\label{h2}
\end{equation}
We see that  that $h(r)$ can be written 
as a sum of two contributions: the pure step function  contribution $-\Theta(D-r)$, 
which has support for $0 \le r <D$, and a contribution
involving $\alpha_2(r;D)$, which has support $D \le r < 2D$.
Substitution of (\ref{h2}) into (\ref{factor2}) yields the structure factor to be
\begin{equation}
S(k)=S_{SF}(k)+ S_{EX}(k),
\label{combined}
\end{equation}
where 
\begin{equation}
S_{SF}(k)=1-2^{\frac{d}{2}}\Gamma(1+d/2)\frac{J_{\frac{d}{2}}(kD)}{(kD)^{\frac{d}{2}}}
\label{SF}
\end{equation}
is the structure factor for the step function  contribution $-\Theta(D-r)$, and
\begin{equation}
S_{EX}(k)=2^{\frac{d}{2}}\Gamma(1+d/2)\int_{D}^{2D}r^{d-1}\frac{\alpha_2(r;D)}{2-\alpha_2(r;D)}
\frac{J_{\frac{d}{2}-1}\!\left(kr\right)}{\left(kr\right)^{\frac{d}{2}-1}}dr 
\label{excess}
\end{equation}
is the contribution to $S(k)$ in excess to the
structure factor for the step function. Here
we have used the fact that the infinite-time density is $\phi=1/2^d$.
For odd dimensions, $S_{EX}(k)$ can be obtained explicitly
in terms of sine, cosine, sine integral and cosine integral functions.
We do not explicitly present these expressions here but instead
plot $S(k)$, defined by (\ref{combined}), for various dimensions in
Fig. \ref{ghosts}. For $d=1,3$ and $11$, $S(k=0)$ is given by
0.150728, 0.290134 and 0.452217, respectively, and therefore
not only is the ghost RSA packing not hyperuniform, as expected,
but its distance from hyperuniformity increases
as the space dimension $d$ increases, asymptotically approaching
the value of 1/2. This should be contrasted with the standard
RSA packing in $\mathbb{R}^d$, which we have shown is nearly hyperuniform.

In the limit $d \rightarrow \infty$, the excess contribution to the structure factor 
has the limiting form
\begin{equation}
S_{EX}(k) \rightarrow
\frac{1}2{}\left[\frac{2^{\frac{d}{2}}\Gamma(1+d/2)J_{\frac{d}{2}}(kD)}{(kD)^{\frac{d}{2}}}\right]^2=
\frac{1}{2}[1-S_{SF}(k)]^2.
\label{asympt}
\end{equation}
The resulting structure factor in this asymptotic limit is given by
\begin{eqnarray}
S(k)&=&S_{SF}(k)+S_{EX}(k) \nonumber \\
&=& \frac{1}{2}[1+S^2_{SF}(k)], \qquad d \rightarrow \infty.
\end{eqnarray}
It has the following small-$k$ expansion:
\begin{equation}
S(k) = \frac{1}{2}+ \frac{1}{8(d+2)^2} k^4 -\frac{1}{16(d+2)^2(d+4)} k^6 + {\cal O}(k^8), \qquad k \rightarrow 0.
\end{equation}
The asymptotic result (\ref{asympt}) is easily obtained by utilizing the
fact that in the limit $d \rightarrow \infty$, $\alpha_2(r;D)/[2-\alpha_2(r;D)] \rightarrow \alpha_2(r;D)/2$ 
\cite{To06b}. Substitution of this result into 
the general relation (\ref{excess}) and recognizing that the lower limit
$D$ of this integral can be replaced by 0 in the limit $d\rightarrow \infty$ yields
the asymptotic form
\begin{equation}
S_{EX}(k) \rightarrow \frac{{\tilde \alpha}_2(k;D)}{2 v_1(D)},
\label{int}
\end{equation}
where ${\tilde \alpha}_2(k;D)$ denotes the Fourier transform of $\alpha_2(r;D)$
and $v_1(D)$ is the volume of a sphere of radius $D$ [cf. (\ref{v(R)})].
The quantity  ${\tilde \alpha}_2(k;D)$ is known explicitly in any dimension
\cite{To03} and substitution of this result into (\ref{int})
immediately yields (\ref{asympt}).

\end{appendix}


\begin{thebibliography}{35}
\expandafter\ifx\csname natexlab\endcsname\relax\def\natexlab#1{#1}\fi
\expandafter\ifx\csname bibnamefont\endcsname\relax
  \def\bibnamefont#1{#1}\fi
\expandafter\ifx\csname bibfnamefont\endcsname\relax
  \def\bibfnamefont#1{#1}\fi
\expandafter\ifx\csname citenamefont\endcsname\relax
  \def\citenamefont#1{#1}\fi
\expandafter\ifx\csname url\endcsname\relax
  \def\url#1{\texttt{#1}}\fi
\expandafter\ifx\csname urlprefix\endcsname\relax\def\urlprefix{URL }\fi
\providecommand{\bibinfo}[2]{#2}
\providecommand{\eprint}[2][]{\url{#2}}

\bibitem[{\citenamefont{Boltzmann}(1898)}]{Bo64}
\bibinfo{author}{\bibfnamefont{L.}~\bibnamefont{Boltzmann}},
  \emph{\bibinfo{title}{Lectures on Gas Theory}}
  (\bibinfo{publisher}{University of California Press},
  \bibinfo{address}{Berkeley, California}, \bibinfo{year}{1898}),
  \bibinfo{note}{1964 translation by S. G. Brush of the original 1898
  publication.}

\bibitem[{\citenamefont{Hansen and McDonald}(1986)}]{Ha86}
\bibinfo{author}{\bibfnamefont{J.~P.} \bibnamefont{Hansen}} \bibnamefont{and}
  \bibinfo{author}{\bibfnamefont{I.~R.} \bibnamefont{McDonald}},
  \emph{\bibinfo{title}{Theory of Simple Liquids}}
  (\bibinfo{publisher}{Academic Press}, \bibinfo{address}{New York},
  \bibinfo{year}{1986}).

\bibitem[{\citenamefont{Torquato}(2002)}]{To02a}
\bibinfo{author}{\bibfnamefont{S.}~\bibnamefont{Torquato}},
  \emph{\bibinfo{title}{Random Heterogeneous Materials: Microstructure and
  Macroscopic Properties}} (\bibinfo{publisher}{Springer-Verlag},
  \bibinfo{address}{New York}, \bibinfo{year}{2002}).


\bibitem[{\citenamefont{Chaikin and Lubensky}(1995)}]{Chaik95}
\bibinfo{author}{\bibfnamefont{P.~M.} \bibnamefont{Chaikin}} \bibnamefont{and}
  \bibinfo{author}{\bibfnamefont{T.~C.} \bibnamefont{Lubensky}},
  \emph{\bibinfo{title}{Principles of Condensed Matter Physics}}
  (\bibinfo{publisher}{Cambridge University Press}, \bibinfo{address}{New
  York}, \bibinfo{year}{1995}).

\bibitem[{\citenamefont{Conway and Sloane}(1998)}]{Co93}
\bibinfo{author}{\bibfnamefont{J.~H.} \bibnamefont{Conway}} \bibnamefont{and}
  \bibinfo{author}{\bibfnamefont{N.~J.~A.} \bibnamefont{Sloane}},
  \emph{\bibinfo{title}{Sphere Packings, Lattices and Groups}}
  (\bibinfo{publisher}{Springer-Verlag}, \bibinfo{address}{New York},
  \bibinfo{year}{1998}).

\bibitem[{\citenamefont{Frisch and Percus}(1999)}]{Fr99}
\bibinfo{author}{\bibfnamefont{H.~L.} \bibnamefont{Frisch}} \bibnamefont{and}
  \bibinfo{author}{\bibfnamefont{J.~K.} \bibnamefont{Percus}},
  \bibinfo{journal}{Phys. Rev. E} \textbf{\bibinfo{volume}{60}},
  \bibinfo{pages}{2942} (\bibinfo{year}{1999}).

\bibitem[{\citenamefont{Parisi and Slanina}(2000)}]{Pa00}
\bibinfo{author}{\bibfnamefont{G.}~\bibnamefont{Parisi}} \bibnamefont{and}
  \bibinfo{author}{\bibfnamefont{F.}~\bibnamefont{Slanina}},
  \bibinfo{journal}{Phys. Rev. E} \textbf{\bibinfo{volume}{62}},
  \bibinfo{pages}{6544} (\bibinfo{year}{2000}).

\bibitem[{\citenamefont{Torquato and Stillinger}(2002)}]{To02c}
\bibinfo{author}{\bibfnamefont{S.}~\bibnamefont{Torquato}} \bibnamefont{and}
\bibinfo{author}{\bibfnamefont{F.~H.} \bibnamefont{Stillinger}},
\bibinfo{journal}{J. Phys. Chem. B} \textbf{\bibinfo{volume}{106}},
\bibinfo{pages}{8354} (\bibinfo{year}{2002}), \bibinfo{note}{{E}rratum {\bf
106}, 11406 {(}2002{)}.}


\bibitem[{\citenamefont{Parisi and Zamponi}(2006)}]{Pa06}
\bibinfo{author}{\bibfnamefont{G.}~\bibnamefont{Parisi}} \bibnamefont{and}
  \bibinfo{author}{\bibfnamefont{F.}~\bibnamefont{Zamponi}},
  \bibinfo{journal}{J. Stat. Mech.} \bibinfo{pages}{P03017}
  (\bibinfo{year}{2006}).

\bibitem[{\citenamefont{Rogers}(1964)}]{Ro64}
\bibinfo{author}{\bibfnamefont{C.~A.} \bibnamefont{Rogers}},
  \emph{\bibinfo{title}{Packing and Covering}} (\bibinfo{publisher}{Cambridge
  University Press}, \bibinfo{address}{Cambridge}, \bibinfo{year}{1964}).

\bibitem[{\citenamefont{Minkowski}(1905)}]{Mi05}
\bibinfo{author}{\bibfnamefont{H.}~\bibnamefont{Minkowski}},
  \bibinfo{journal}{J. {r}eine {a}ngew. {M}ath.}
  \textbf{\bibinfo{volume}{129}}, \bibinfo{pages}{220} (\bibinfo{year}{1905}).

\bibitem[{\citenamefont{Kabatiansky and Levenshtein}(1978)}]{Ka78}
\bibinfo{author}{\bibfnamefont{G.~A.} \bibnamefont{Kabatiansky}}
  \bibnamefont{and} \bibinfo{author}{\bibfnamefont{V.~I.}
  \bibnamefont{Levenshtein}}, \bibinfo{journal}{Problems of Information
  Transmission} \textbf{\bibinfo{volume}{14}}, \bibinfo{pages}{1}
  (\bibinfo{year}{1978}).

\bibitem[{\citenamefont{Torquato and Stillinger}(2006{\natexlab{a}})}]{To06a}
\bibinfo{author}{\bibfnamefont{S.}~\bibnamefont{Torquato}} \bibnamefont{and}
  \bibinfo{author}{\bibfnamefont{F.~H.} \bibnamefont{Stillinger}},
  \bibinfo{journal}{Phys. Rev. E} \textbf{\bibinfo{volume}{73}},
  \bibinfo{pages}{031106} (\bibinfo{year}{2006}{\natexlab{a}}).

\bibitem[{\citenamefont{Torquato and Stillinger}(2006{\natexlab{b}})}]{To06b}
\bibinfo{author}{\bibfnamefont{S.}~\bibnamefont{Torquato}} \bibnamefont{and}
  \bibinfo{author}{\bibfnamefont{F.~H.} \bibnamefont{Stillinger}},
  \bibinfo{journal}{Experimental Math.}  (\bibinfo{year}{2006}{\natexlab{b}}),
  \bibinfo{note}{in press; see also arXiv:math.MG/0508381.}

\bibitem{Ta00}
 J. Talbot, G. Tarjus, P. R.
 Van Tassel, and P. Viot, Colloids and Surfaces A, {\bf 165}, 287 (2000).
 

\bibitem{footnote1}
Note that the infinite-time RSA limit has also been called the ``jammed" limit,
but, for general packing problems, the term {\it jammed} has more recently come
to signify a certain degree of mechanical rigidity due to 
the existence of an interparticle contact network. The reader interested
in the latter sense of jamming is referred to 
S. Torquato and F.~H. Stillinger, J. Phys. Chem. B {\bf 105}, 11849 (2001);
S. Torquato, A. Donev, and F.~H. Stillinger, Int. J. Solids Structures {\bf 40},
7143 (2003); and  A. Donev, S. Torquato, F.~H. Stillinger and R. Connelly,
J. Comp. Phys. {\bf 197}, 139 (2004).

\bibitem[{\citenamefont{Re{\'n}yi}(1963)}]{Re63}
\bibinfo{author}{\bibfnamefont{A.}~\bibnamefont{Re{\'n}yi}},
  \bibinfo{journal}{Sel. Trans. Math. Stat. Prob.}
    \textbf{\bibinfo{volume}{4}}, \bibinfo{pages}{203} (\bibinfo{year}{1963}).


\bibitem[{\citenamefont{Feder}(1980)}]{Fe80}
\bibinfo{author}{\bibfnamefont{J.}~\bibnamefont{Feder}}, \bibinfo{journal}{J.
  Theor. Biol.} \textbf{\bibinfo{volume}{87}}, \bibinfo{pages}{237}
  (\bibinfo{year}{1980}).

\bibitem{footnote2} In Ref. \cite{To06b}, a {\it disordered packing} in $\Re^d$
is defined to be a packing in which the pair correlation function $g_2({\bf r})$ decays
to its long-range value of unity faster than $|{\bf r}|^{-d-\epsilon}$ for
some $\epsilon >0$.


\bibitem[{\citenamefont{Hinrichsen et~al.}(1986)\citenamefont{Hinrichsen,
  Feder, and Jossang}}]{Hi86}
\bibinfo{author}{\bibfnamefont{E.}~\bibnamefont{Hinrichsen}},
  \bibinfo{author}{\bibfnamefont{J.}~\bibnamefont{Feder}}, \bibnamefont{and}
  \bibinfo{author}{\bibfnamefont{T.}~\bibnamefont{Jossang}},
  \bibinfo{journal}{J. Stat. Phys.} \textbf{\bibinfo{volume}{44}},
  \bibinfo{pages}{793} (\bibinfo{year}{1986}).

\bibitem[{\citenamefont{Cooper}(1988)}]{Co88}
\bibinfo{author}{\bibfnamefont{D.~W.} \bibnamefont{Cooper}},
  \bibinfo{journal}{Phys. Rev. A} \textbf{\bibinfo{volume}{38}},
  \bibinfo{pages}{522} (\bibinfo{year}{1988}).

\bibitem{Ta91}
J. Talbot, P. Schaaf and G. Tarjus, Mol. Phys., {\bf 72}, L397 (1991).

\bibitem{To03}
S. Torquato and F. H. Stillinger,
Phys. Rev. E, {\bf 68}, 041113 1 (2003);  {\bf 68}, 069901 (2003).

\bibitem[{\citenamefont{Pomeau}(1980)}]{Po80}
\bibinfo{author}{\bibfnamefont{Y.}~\bibnamefont{Pomeau}}, \bibinfo{journal}{J.
  Phys. A: Math. Gen.} \textbf{\bibinfo{volume}{13}}, \bibinfo{pages}{L193}
  (\bibinfo{year}{1980});
  
\bibitem{Sw81}  
\bibinfo{author}{\bibfnamefont{R.~H.} \bibnamefont{Swendsen}},
  \bibinfo{journal}{Phys. Rev. A} \textbf{\bibinfo{volume}{24}},
  \bibinfo{pages}{504} (\bibinfo{year}{1981}).

\bibitem[{\citenamefont{Bonnier et~al.}(1994)\citenamefont{Bonnier, Boyer, and
  Viot}}]{Bo94}
\bibinfo{author}{\bibfnamefont{B.}~\bibnamefont{Bonnier}},
  \bibinfo{author}{\bibfnamefont{D.}~\bibnamefont{Boyer}}, \bibnamefont{and}
  \bibinfo{author}{\bibfnamefont{P.}~\bibnamefont{Viot}}, \bibinfo{journal}{J.
  Phys. A: Math. Gen.} \textbf{\bibinfo{volume}{27}}, \bibinfo{pages}{3671}
  (\bibinfo{year}{1994}).

\bibitem[{\citenamefont{Brosilow et~al.}(1991)\citenamefont{Brosilow, Ziff, and
Vigil}}]{Bro91}
\bibinfo{author}{\bibfnamefont{B.~J.} \bibnamefont{Brosilow}},
\bibinfo{author}{\bibfnamefont{R.~M.} \bibnamefont{Ziff}}, \bibnamefont{and}
\bibinfo{author}{\bibfnamefont{R.~D.} \bibnamefont{Vigil}},
\bibinfo{journal}{Phys. Rev. A} \textbf{\bibinfo{volume}{43}},
\bibinfo{pages}{631} (\bibinfo{year}{1991}).


\bibitem{footnote3}
This time-saving idea is similar to that used in Ref. \cite{Bro91}
for oriented squares in two dimensions but its implementation 
for hyperspheres is necessarily different.



\bibitem{Ba92}
K.~Ball, Int. Math. Res. Notices {\bf 68} 217 (1992).


\bibitem[{\citenamefont{Pal{\'a}sti}(1960)}]{Pal60}
\bibinfo{author}{\bibfnamefont{I.}~\bibnamefont{Pal{\'a}sti}},
  \bibinfo{journal}{Publ. Math. Inst. Hung. Acad. Sci.}
  \textbf{\bibinfo{volume}{5}}, \bibinfo{pages}{353} (\bibinfo{year}{1960}).


\bibitem[{\citenamefont{Torquato et~al.}(2000)\citenamefont{Torquato, Truskett,
and Debenedetti}}]{To00b}
\bibinfo{author}{\bibfnamefont{S.}~\bibnamefont{Torquato}},
\bibinfo{author}{\bibfnamefont{T.~M.} \bibnamefont{Truskett}},
\bibnamefont{and} \bibinfo{author}{\bibfnamefont{P.~G.}
\bibnamefont{Debenedetti}}, \bibinfo{journal}{Phys. Rev. Lett.}
\textbf{\bibinfo{volume}{84}}, \bibinfo{pages}{2064} (\bibinfo{year}{2000}).

\bibitem[{\citenamefont{{Kansal} et~al.}(2002)\citenamefont{{Kansal},
{Torquato}, and {Stillinger}}}]{Ka02}
\bibinfo{author}{\bibfnamefont{A.~R.} \bibnamefont{{Kansal}}},
\bibinfo{author}{\bibfnamefont{S.}~\bibnamefont{{Torquato}}},
\bibnamefont{and} \bibinfo{author}{\bibfnamefont{F.~H.}
\bibnamefont{{Stillinger}}}, \bibinfo{journal}{Phys. Rev. E}
\textbf{\bibinfo{volume}{66}}, \bibinfo{pages}{041109}
(\bibinfo{year}{2002}).

\bibitem{Do05d}
 A. Donev, F. H. Stillinger, and S. Torquato, Phys. Rev. Lett. {\bf 95}, 090604 (2005).

 \bibitem{Sk06}
 M. Skoge, A. Donev, F. H. Stillinger and S. Torquato, Phys. Rev. E, in press.



\bibitem[{\citenamefont{Costin and Lebowitz}(2004)}]{Cos04}
\bibinfo{author}{\bibfnamefont{O.}~\bibnamefont{Costin}} \bibnamefont{and}
  \bibinfo{author}{\bibfnamefont{J.}~\bibnamefont{Lebowitz}},
  \bibinfo{journal}{J. Phys. Chem. B.} \textbf{\bibinfo{volume}{108}},
  \bibinfo{pages}{19614} (\bibinfo{year}{2004}).


\bibitem{Ya61}
M.~Yamada, Prog. Theor. Phys. {\bf 25}, 579 (1961).

\bibitem[{\citenamefont{Crawford et~al.}(2003)\citenamefont{Crawford, Torquato,
  and Stillinger}}]{Cr03}
\bibinfo{author}{\bibfnamefont{J.~R.} \bibnamefont{Crawford}},
  \bibinfo{author}{\bibfnamefont{S.}~\bibnamefont{Torquato}}, \bibnamefont{and}
  \bibinfo{author}{\bibfnamefont{F.~H.} \bibnamefont{Stillinger}},
  \bibinfo{journal}{J. Chem. Phys.} \textbf{\bibinfo{volume}{119}},
  \bibinfo{pages}{7065} (\bibinfo{year}{2003}).

\bibitem{Uc06a}
O.~U. Uche, F. H. Stillinger and S. Torquato, 
Physica A {\bf 360}, 21 (2006).
 


\bibitem[{\citenamefont{Donev et~al.}(2005)\citenamefont{Donev, Torquato, and
  Stillinger}}]{Do05c}
\bibinfo{author}{\bibfnamefont{A.}~\bibnamefont{Donev}},
  \bibinfo{author}{\bibfnamefont{S.}~\bibnamefont{Torquato}}, \bibnamefont{and}
  \bibinfo{author}{\bibfnamefont{F.~H.} \bibnamefont{Stillinger}},
  \bibinfo{journal}{Phys. Rev. E} \textbf{\bibinfo{volume}{71}},
  \bibinfo{pages}{011105} (\bibinfo{year}{2005}).

\bibitem[{\citenamefont{{Hales}}(2005)}]{Ha05}
\bibinfo{author}{\bibfnamefont{T.~C.} \bibnamefont{{Hales}}},
\bibinfo{journal}{Ann. Math.} \textbf{\bibinfo{volume}{162}},
\bibinfo{pages}{1065} (\bibinfo{year}{2005}).
      

\bibitem[{\citenamefont{Musin}()}]{Mu04}
\bibinfo{author}{\bibfnamefont{O.}~\bibnamefont{Musin}},
\bibinfo{note}{Technical Report, Moscow State University, 2004}.
  

\end{thebibliography}

\end{document}